\documentstyle[aps,epsfig]{revtex}

\begin{document}

\title{On the properties of small-world network models}

\author{A. Barrat$^1$ and M. Weigt$^2$}

\address{
$^1$ Laboratoire de Physique Th{\'e}orique
\cite{umr}, B{\^a}timent 210, Universit{\'e}  
de Paris-Sud, 91405 Orsay Cedex, France ; email: 
Alain.Barrat@th.u-psud.fr \\
$^2$ CNRS-Laboratoire de Physique Th{\'e}orique de l'E.N.S., 24 rue Lhomond,
75231 Paris Cedex 05, France; email: weigt@physique.ens.fr }

\date{\today}

\maketitle
\begin{abstract}
We study the small-world networks recently introduced by Watts and
Strogatz [Nature {\bf 393}, 440 (1998)], using analytical as well as 
numerical tools. We characterize the geometrical properties resulting
from the coexistence of a local structure and random long-range
connections, and we examine their evolution with size and disorder 
strength. We show that any finite value of 
the disorder is able to trigger a ``small-world'' behaviour as soon as
the initial lattice is big enough, and study the crossover between a 
regular lattice and a ``small-world'' one. These results are
corroborated by the investigation of an Ising model
defined on the network, showing for every finite disorder fraction 
a crossover from a high-temperature region
dominated by the underlying one-dimensional structure to a mean-field
like low-temperature region. In particular there exists a
finite-temperature ferromagnetic phase
transition as soon as the disorder strength is finite.\\[0.5cm]
\end{abstract}

%\PACS{
%{05.50.+q}{Lattice theory and statistics}  \and
%{64.60.C}{Order-disorder transformations and statistical mechanics
%of model systems}  \and
%{05.70.Fh}{Phase transitions: general studies}
%}
%
%\maketitle

         % of model systems)
05.70.Fh %(Phase transitions: general studies)

\section{Introduction}

A recent article by Watts and Strogatz \cite{watts_strogatz},
showing the relevance of what they called 
``small-world'' networks for many realistic situations,
has triggered a lot of attention for these kind of networks
\cite{bart,comment,remi,newman,argollo,newman2}:
this interest results
from their very definition, allowing an exploration between
regular and random networks.

Random networks have of course been the subject of many studies in
various domains, ranging from physics to social sciences.
A very important characteristic common to such lattices
and for example social networks is that the length of the shortest
chain connecting two vertices (or members) grows very slowly,
i.e. in general logarithmically,
with the size of the network \cite{bollobas}. This characteristic
has important consequences for many issues, e.g. the speed of
disease spreading \cite{watts_strogatz} etc. The social psychologist 
S. Milgram \cite{milgram}, after realizing that the number of persons
necessary to link two randomly chosen, geographically separated persons
had a median number of six, has called this concept the ``six
degrees of separation''. In addition, models defined on
random networks are, due to their locally tree-like structure,
of mean-field type, and can therefore be analytically more tractable
than their counterparts defined on regular lattices, but, thanks to the
finite connectivity of their vertices, they
display however behaviours which are intrinsically not captured
by the familiar infinite connectivity models \cite{barrat}.

However, it is well known that many realistic networks have a local
structure which is very different from random networks with finite
connectivity. For example, two neighbours have many
common neighbours, a property which does not hold for random networks,
and which can be quantified by the introduction of the
``clustering coefficient'' (see section \ref{geom}). Such phenomena are not
only found in social networks, but also e.g. in the connections
of neural networks \cite{watts_strogatz} or in the chemical bond
structure of long macromolecules \cite{degennes}: The one-dimensional
couplings of neighbouring monomers are complemented by long-ranged
interactions between monomers that are close in space although not
along the chain. This interplay has been studied
in fact for example in \cite{chakra}, but it seems that, in this case,
the long-range interactions are not sufficient to really modify
the properties of the one-dimensional structure of the chain
\footnote{for example, an Ising model defined on a self-avoiding 
walk with interactions between monomers neighbours in space and not only
on the chain has a critical temperature $T_c=0$, as for a one-dimensional
chain \cite{chakra}}.

The construction proposed by Watts and Strogatz \cite{watts_strogatz},
that we will recall in section \ref{def}, allows to reconcile local properties
of a regular network with global properties of a random one, by 
introducing a certain amount of random long-range connections into an 
initially regular network. 

The aim of this paper is to study in some detail the concepts
used in \cite{watts_strogatz} to characterize the ``small-world'' behaviour,
caused by the coexistence of ``short-range'' and ``long-range''
connections.
We will show that this behaviour does not appear at a finite value of
the disorder $p$, but that, for any $p>0$, the networks will display
this behaviour as soon as their size is large enough. 

This paper is organized as follows.
In section \ref{def}
we describe the procedure used to obtain small-world networks;
in section \ref{geom} we study some of their geometrical
properties, i.e. the connectivity,
the chemical distances and the ``clustering'' coefficient, analytically
as well as numerically
\footnote{Results in particular about the chemical distances
and the onset of the small-world behaviour can also be found in 
\cite{bart,comment,newman,argollo,newman2}}.
Section IV contains the investigation of an
Ising-model defined on a small-world lattice, where the interplay between the
short- and long-range interactions leads to interesting physical effects.

\section{Definition of the model(s)}
\label{def}

The construction algorithm proposed by Watts and Strogatz for
small-world networks is the following:
the initial network is a one-dimensional lattice of $N$ sites, with periodic
boundary conditions (i.e. a ring), each vertex being connected to its
$2k$ nearest neighbours. The vertices are then
visited one after the other;
each link connecting a vertex to one of its
$k$ nearest neighbours in the clockwise sense
is left in place with probability $1-p$, and
with probability $p$ is reconnected to a randomly chosen other vertex.  
Long range connections are therefore introduced.
Note that, even for $p=1$, the network keeps some memory of the procedure
and is not locally equivalent to a random network: each vertex has
indeed {\it at least} $k$ neighbours. An important consequence is that
we have no isolated vertices, and the graph has usually only one component
(a random graph has usually many components of various sizes).

\begin{figure}[bt]
\centerline{
        \epsfig{figure=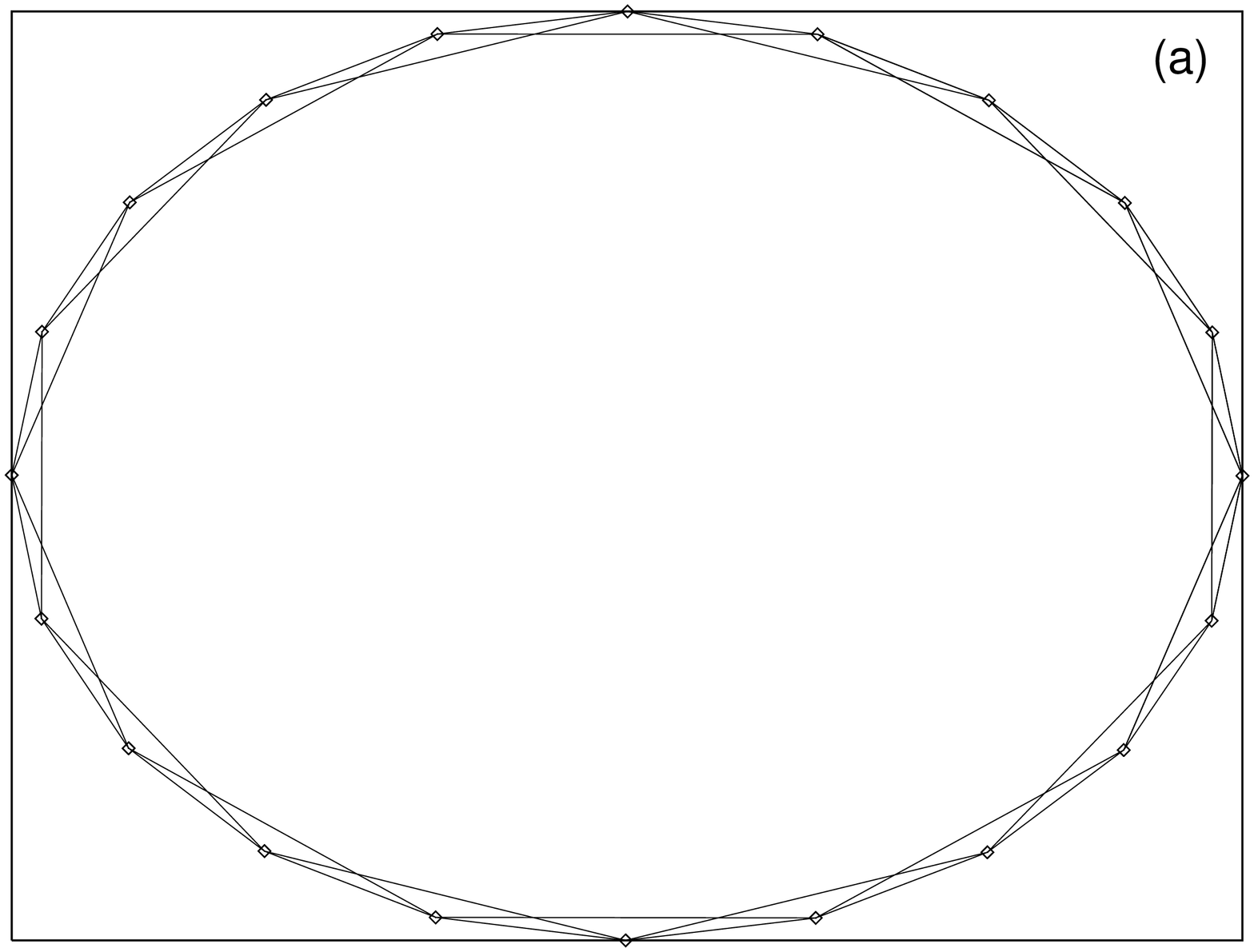,width=4cm,angle=-90}
        \epsfig{figure=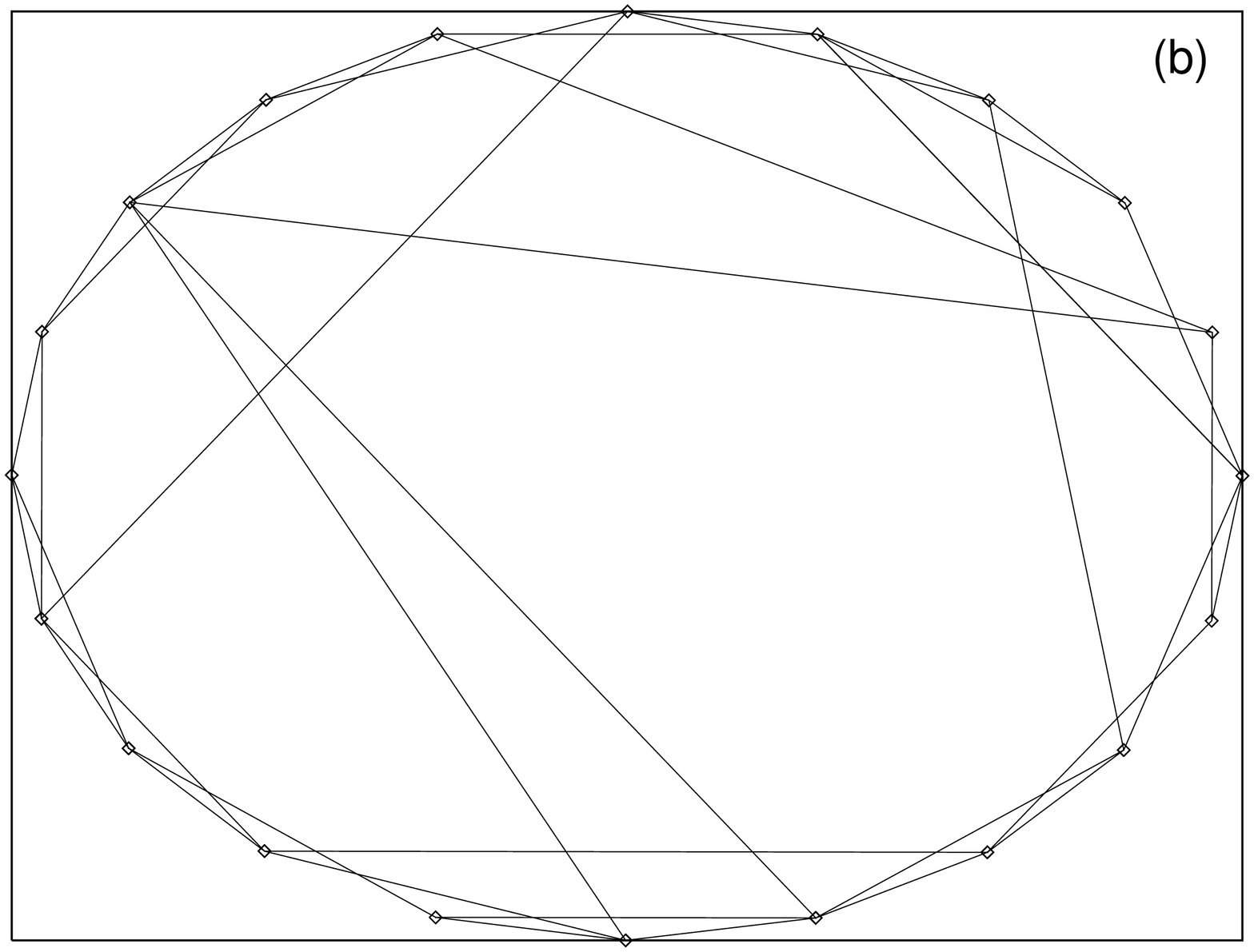,width=4cm,angle=-90}
        }
\centerline{
        \epsfig{figure=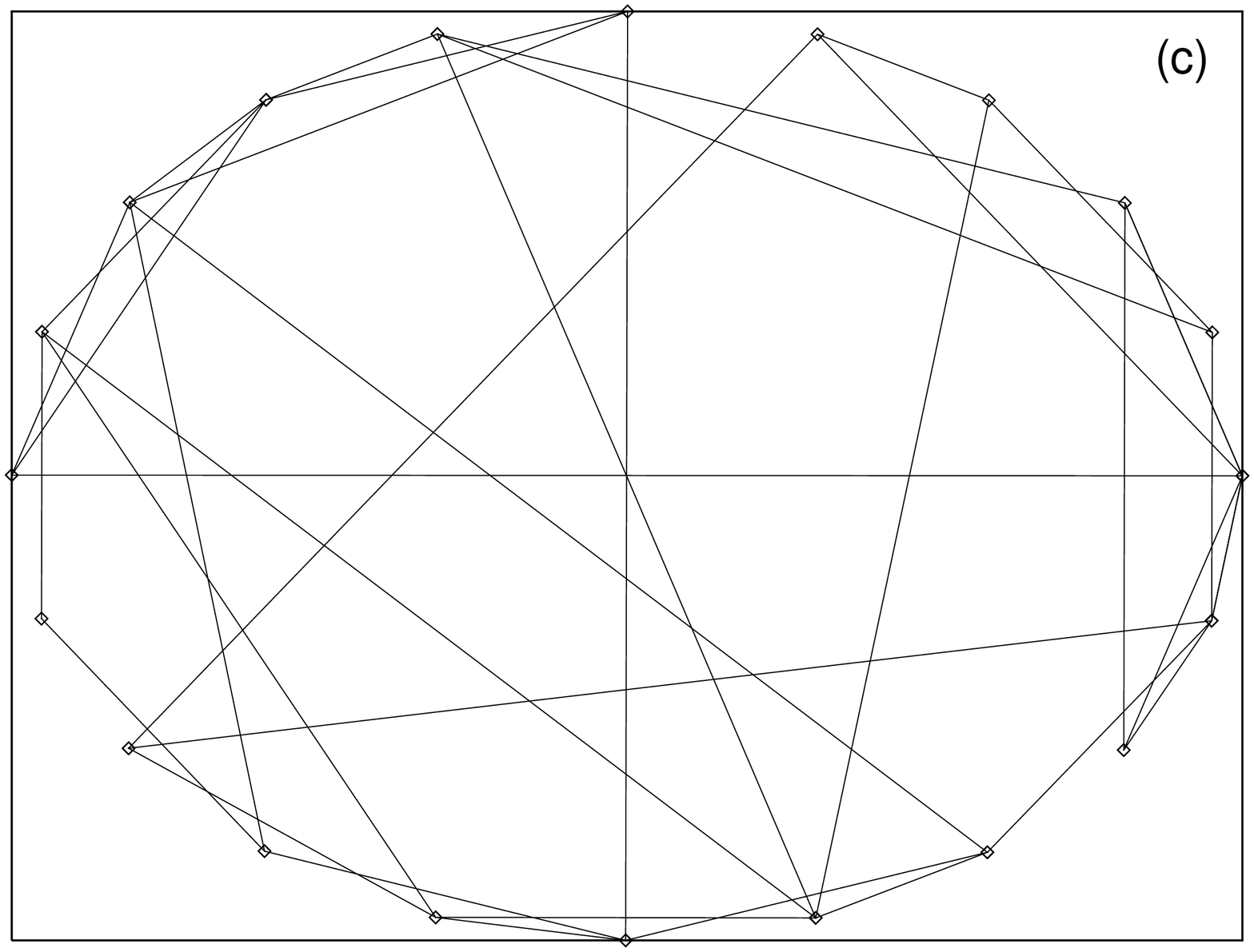,width=4cm,angle=-90}
        \epsfig{figure=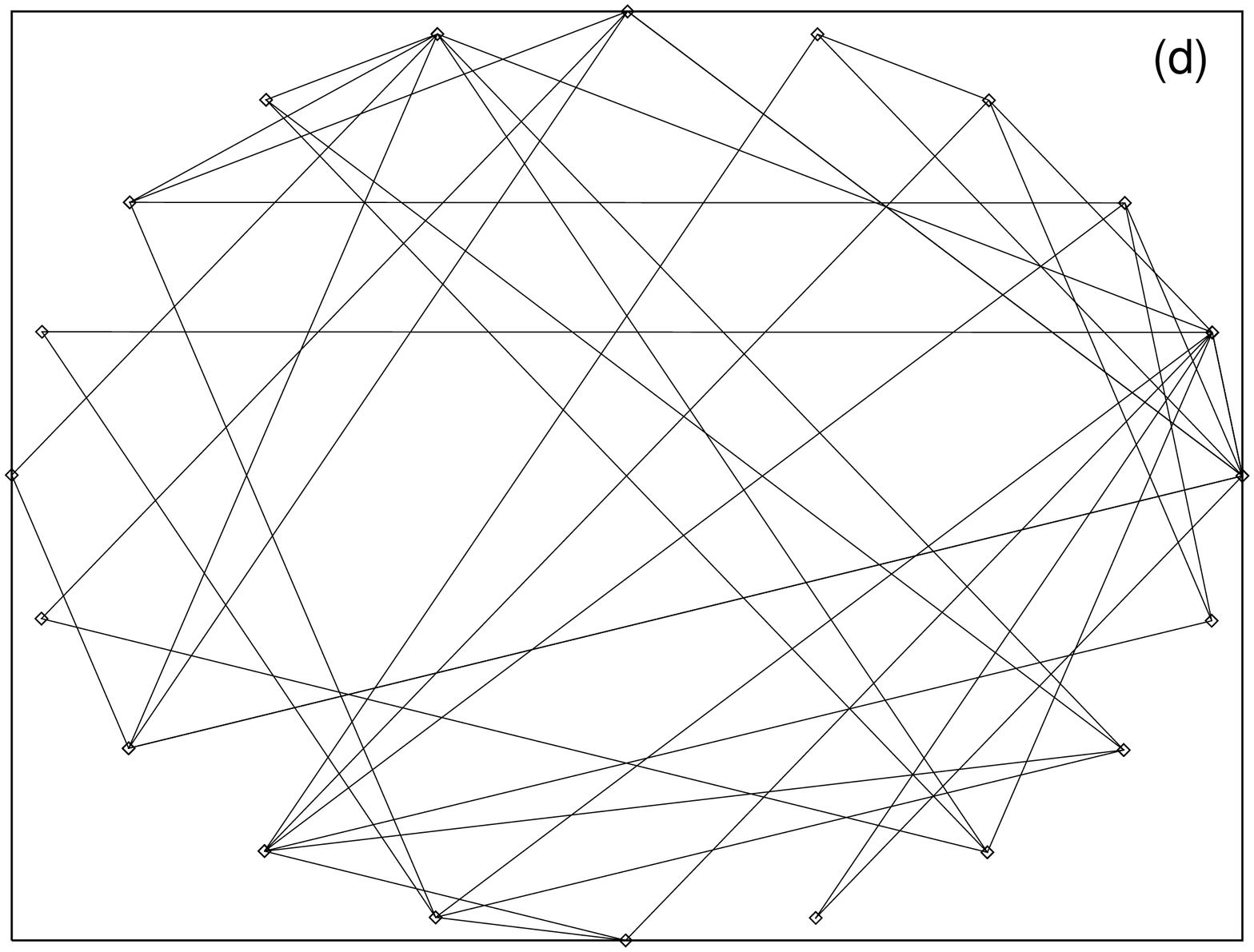,width=4cm,angle=-90}
        }
\caption{Examples of networks obtained by the procedure described in the text,
for $k=2$, $N=20$.  (a): $p=0$, regular networks; (b), (c):
intermediate values of $p$; (d): $p=1$.}
\label{fig:network}
\end{figure}

It is possible to obtain ``small-world'' networks 
in other ways, that yields the same physical consequences, and
can be more tractable analytically. For example, the
networks studied in \cite{remi,newman}
are obtained by {\it adding} long-range connections
to the initial ring without diluting its one-dimensional structure; 
the mean connectivity then changes with the
disorder. In section \ref{Ising} we will also study an initial network
with multiple links between successive vertices.

\section{Geometrical properties}
\label{geom}

\subsection{Connectivity}
\label{geomA}

For $p=0$, each vertex has the same connectivity $2k$. On the other hand, a 
non-zero value of $p$ introduces disorder into the network, in the form of
a non-uniform connectivity, while maintaining a fixed average
connectivity $\bar{c} = 2k$. Let us denote $P_p(c)$ the probability
distribution of the connectivities.

\begin{figure}[bt]
\centerline{
        \epsfig{figure=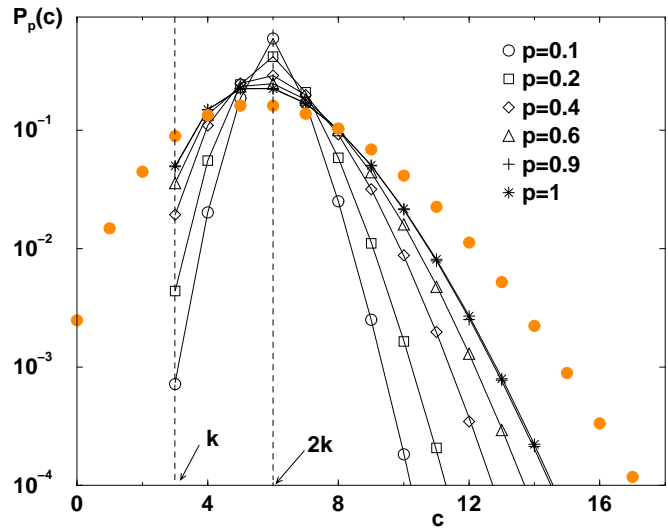,width=7cm,angle=-90}
        }
\caption{Probability distributions of the connectivity $c$
for $k=3$ and various values of $p$:
$c \ge k$, and the mean connectivity is $\bar{c} = 2k=6$.
The symbols are obtained by numerical simulations of small-world
networks (with $N=1000$ vertices),
and the lines are a guide to the eye, joining points
given by formula (\ref{eq:prob}). Filled circles show the
probability distribution of the connectivity $c$ for a random network
of mean connectivity is $\bar{c} = 2k=6$ (given by $(2k)^c \exp(-2k)/c!$).}
\label{fig:connect}
\end{figure}

Since $k$ of the initial $2k$ connections of each vertex are left untouched
by the construction, the connectivity of a vertex $i$ can be written
$c_i=k+n_i$, with $n_i \ge 0$. $n_i$ can then again be divided in two parts:
$n_i^1 \le k$ links have been left in place (each one with probability $1-p$),
the other $n_i^2=n_i-n_i^1$ links have been reconnected {\it towards}
$i$, each one with probability $p/N$. We readily obtain
\begin{equation}
P_1(n_i^1)= {k \choose n_i^1} (1-p)^{n_i^1} p^{k-n_i^1}
\end{equation}
\begin{equation}
P_2(n_i^2) = \frac{(k p)^{n_i^2} }{n_i^2 !} \exp{(-p k)} 
\ \ \ \ \ \ \mbox{  for large $N$}
\end{equation}
and find
\begin{equation}
P_p(c) = \sum_{n=0}^{\min (c-k,k)}
{k \choose n}  (1-p)^{n} p^{k-n}
\frac{(k p)^{c-k-n}}{(c-k-n)!} 
\exp{(-p k)}, \ \ \  c \ge k  .
\label{eq:prob}
\end{equation}
We show in figure (\ref{fig:connect})
the probability distributions for $k=3$ and various values of $p$: as
$p$ grows, the distribution becomes broader.

\subsection{Chemical distances}
\label{geomB}

We now turn to a non-local quantity of graphs:
the chemical distance 
between its vertices, i.e. the minimal number of links
between two vertices. We note $d_{ij}$ the chemical distance between
vertices $i$ and $j$, and
\begin{equation}
\ell(N,p) = \frac{1}{N(N-1)} \overline{\sum_{i \neq j} d_{ij}}
\end{equation}
the mean chemical distance, averaged over all pairs of vertices and
over the disorder induced by the rewiring procedure.

\begin{figure}[bt]
\centerline{
        \epsfig{figure=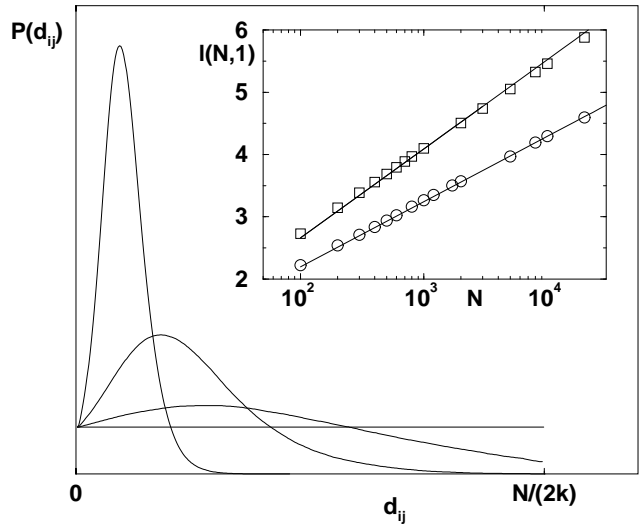,width=7cm,angle=-90}
        }
\caption{Probability distribution of the distance $d_{ij}$ between
two vertices $i$ and $j$ of small-world graphs, for $k=3$, $N=2000$,
$p=2^{-20}$ (flat distribution), $p=2^{-12},\ 2^{-10}$ and
$2^{-8}$ (curves becoming more and more peaked as $p$ grows), averaged
over $500$ samples for each $p$. The maximum value of
$d_{ij}$ is of course $N/(2k)$. Inset:
$\ell(N,1)$ versus $N$ for $k=3$ and $k=5$, together
with the $\ln(N)/\ln(2k-1)$ straight lines.}
\label{fig:distrlength}
\end{figure}

Watts and Strogatz have shown that the mean distance 
between vertices $\ell(N,p)$ decreases very rapidly 
as soon as $p$ is non-zero. They however show the curve
of $\ell(N,p)$ versus $p$ for only one value of $N$ and do not
study how it depends on $N$. 
For $p=0$, we have a linear chain of sites, so that we easily find
\begin{equation}
\ell(N,0) = \frac{ N(N+2k-2)}{ 4k (N-1)} \sim \frac{N}{4k} ,
\label{eq:l0}
\end{equation}
growing like $N$.
On the other hand, for $p=1$ $\ell(N,1)$ grows like 
$\ln (N)/\ln(2k-1)$ (inset of figure (\ref{fig:distrlength})): 
the graph is then random.
Besides, the distribution of lengths, being uniform between 
$1$ (shortest possible distance) and $N/(2k)$ for the linear chain,
becomes more and more peaked around its mean value as $p$ grows
(see figure (\ref{fig:distrlength})).

It is therefore quite natural to ask if the change between
these two behaviours occurs by a transition at a certain finite critical
value of $p$ or if there is a crossover phenomenon at any finite value
of $N$, with a transition occurring only at $p=0$. This last scenario
was first proposed in \cite{bart}.

We first investigate this question by numerical simulations,
to study the behaviour of $\ell(N,p)$ in a systematic way,
varying $N$ and $p$:
we use values of $N$ from $100$ to $20000$, with $p=2^a/2^{20}$,
$a=0,\cdots,20$, and we average over $500$ realizations of the
disorder for each value of $p$. We have studied three different
values of the mean connectivity: $2k=4,\ 6$ and $10$.

\begin{figure}[bt]
\centerline{
        \epsfig{figure=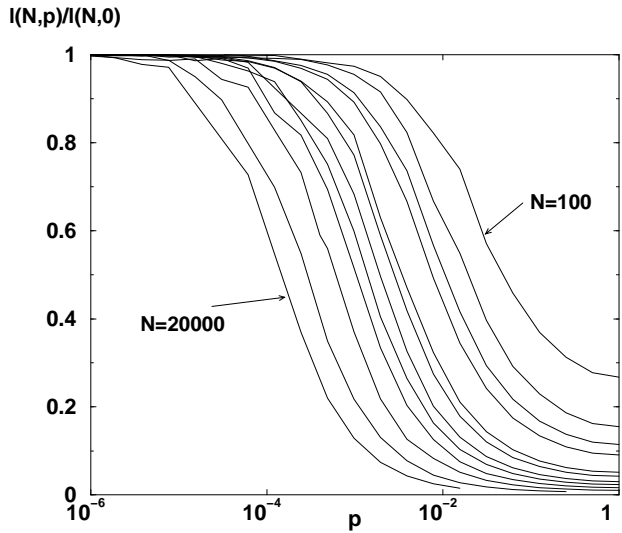,width=7cm,angle=-90}
        }
\caption{Mean chemical length $\ell(N,p)$ normalized by $\ell(N,0)$,
versus $p$, for $k=2$, and $N$ from $100$ to $20000$: the drop in the curve
occurs at lower and lower values of $p$ as $N$ grows.}
\label{fig:lmoy}
\end{figure}

In figure (\ref{fig:lmoy}), we plot $\ell(N,p)/\ell(N,0)$ for various
values of $N$ and $k=2$. It is clear that
$\ell(N,p)$ decreases very fast already for small $p$ (note
the logarithmic scale for $p$): from this point of view,
the network is very soon similar to a random network.
In particular, as $N$ becomes larger, the drop in the curve occurs for
smaller and smaller values of $p$, showing that no finite critical value of
$p$ can be determined this way: in the thermodynamic limit,
$\ell(N,p)/\ell(N,0)$ goes to $0$ for all $p > 0$.
This is a first clear indication of a crossover behaviour 
(as opposed to a transition at a non-zero $p$) that we are now
going to examine in more details.

Note that
the first evidence of a crossover has been given in \cite{bart}
by the numerical study of system with sizes up to $N=1000$, and mean 
connectivities $2k=10,\ 20,\ 30$. 
A scaling of the form
\begin{equation}
\ell(N,p) \sim N^* F_k \left( \frac{N}{N^*} \right)
\label{ansatz}
\end{equation}
was proposed, where $F_k$ depends only on $k$, with
$F_k(u \ll  1) \sim u$, $F_k( u \gg  1) \sim \ln u$, and 
$N^* \sim p^{-\tau}$ with $\tau = 2/3$ as $p$ goes to zero.
However, it can be shown \cite{comment}, with a simple but
rigorous argument, that $\tau$ cannot
in fact be lower than $1$: the mean number of rewired links
is $N_r = p N k$; if $\tau < 1$, and if we take $\alpha$ such that
$\tau < \alpha < 1$, then the scaling hypothesis implies, for large $N$,
$\ell(N,N^{-1/\alpha})
\sim N^{\tau/\alpha} \ln (N)$ (since $N^{1-\tau/\alpha} \gg 1$ for
large $N$); $N_r$ however goes to zero for large $N$, so that the rewiring
of a vanishing number of links could lead to a change in the scaling of 
$\ell$. This obviously unphysical result shows that the hypothesis 
$\tau < 1$ is not valid.
% \footnote{the argument of \cite{comment}
%uses the fact that the mean number of rewired links is $N_r = p N k$,
%and that, if $\tau < 1$, it would be possible to change
%the scaling of $\ell$ (from $\ell \sim N$ 
%to $\ell \sim N^{\tau/\alpha} \ln (N)$, with $\tau < \alpha < 1$),
%with a vanishing {\it number} of rewired links
%($N_r = N^{1-1/\alpha} k $), which is obviously
%unphysical.  }.
In addition, Newman and Watts \cite{newman},
using a renormalization group analysis, have shown that $\tau=1$
exactly.
Here we will arrive at the same result, using our numerical
simulations to test the scaling hypothesis,
as well as analytical arguments.

To understand how strong the disorder has to be to induce a crossover,
and to show that this crossover can occur, at fixed $p$,
for $N^* \sim p^{-\tau}$, or equivalently, at fixed
$N$, for $p^*  \sim N^{-1/\tau}$, only with $\tau \geq 1$,
we study the case of a finite number of rewired links,
$N_r=\alpha$. This corresponds to $p= \alpha /N$.
In order to show that such a value of $p$ is not able to alter
the scaling of $\ell$ with $N$, we now establish a
rigorous lower bound.

For any given sample, the extremities of the $\alpha$ rewired links
determine $2\alpha$ intervals. The sum of their lengths on the ring
is $N$, so that at least one of them has a length
of order $N$, which is, even more precisely, larger than $N/(2\alpha)$.
We call this interval $J=[i_0,j_0]$ and we
consider the interval $I \subset J$, of length 
$N/(4\alpha) =bN$,  $I=[i_0+N/(8\alpha),i_0+3N/(8\alpha)]$, which
has not been modified by the rewiring procedure. 
We now decompose the mean length between two vertices of the sample,
$$
\ell  = \frac{1}{N(N-1)} \sum_{i \neq j} d_{ij},
$$
into two contributions: the first one comes from the pairs
$(i,j)$ with $i \in I,\ j \in I$, the second one includes all pairs
$(i,j)$ where at least one of the vertices is not an element of $I$. 
The first contribution can be estimated by formula (\ref{eq:l0}),
since it comes from a part of the graph
which has not been modified, and at a distance big enough
from any modified link:
$$
\sum_{i\in I,\ j \in I} d_{ij} \geq (bN)(bN-1) \frac{bN}{4k} 
$$
(the inequality comes from the fact that we do not have periodic 
boundary conditions for this interval).
We now have access to a lower bound of $\ell (N,\alpha/N)$
(which  is valid for any sample, and consequently also 
for the average over samples):
$$
\ell (N,\alpha/N) \geq \frac{1}{N(N-1)} \sum_{i\in I,\ j \in I} d_{ij} 
\geq \frac{b^3}{4k} N .
$$
Since $\ell (N,\alpha/N)$ is smaller than $\ell(N,0) \sim N/(4k)$,
this shows that
\begin{equation}
\ell (N,\alpha/N) = O(N)   .
\end{equation}
In other words, a {\it finite} number of rewired links 
{\it cannot} change the scaling at large $N$: $\ell (N,\alpha/N)$
is of order $N$ for any finite $\alpha$.

\begin{figure}[bt]
\centerline{
        \epsfig{figure=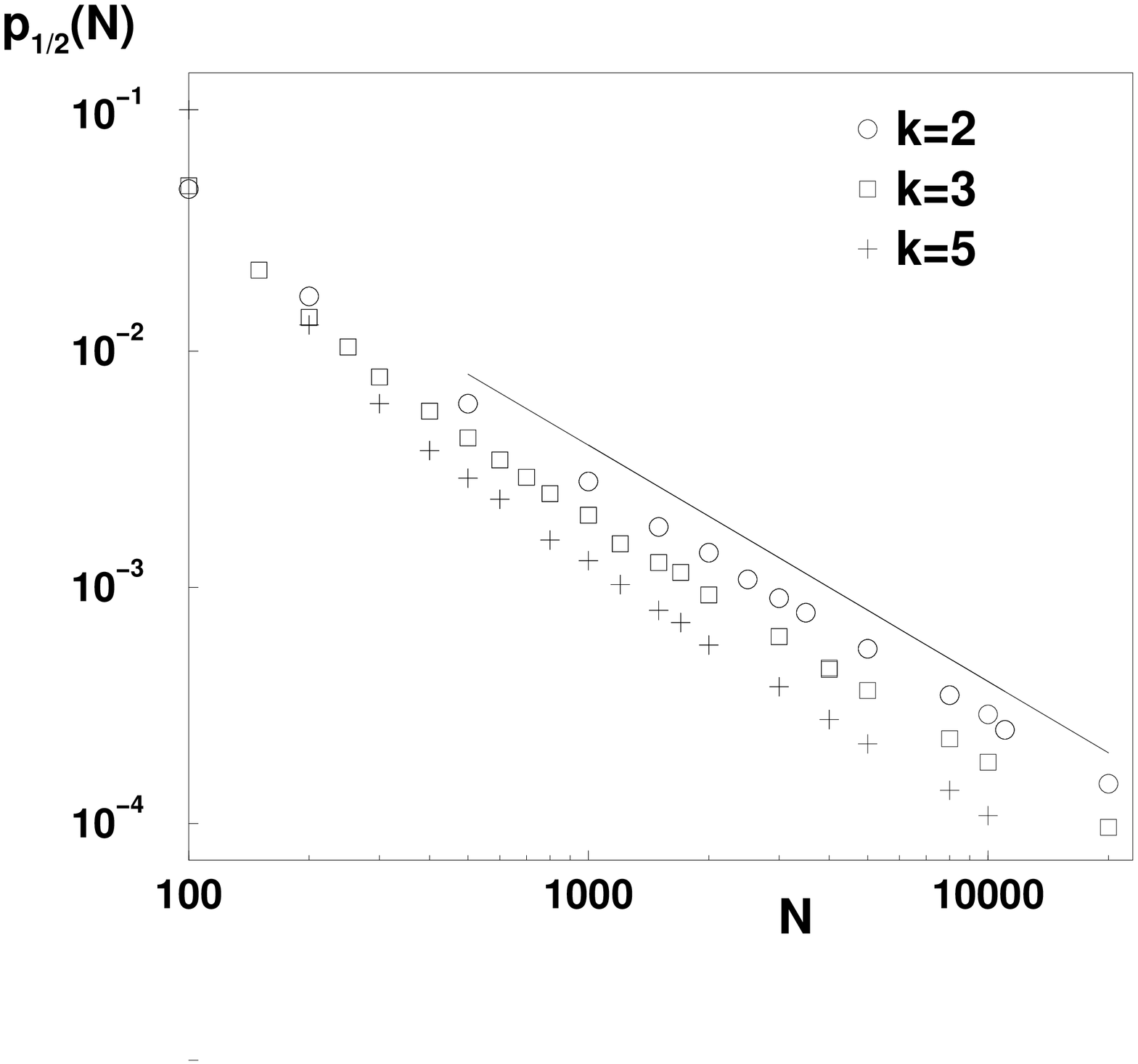,width=7cm,angle=-90}
        }
\caption{$p_{1/2} (N)$ such that
$\ell(N,p_{1/2}(N)) = \ell(N,0) /2 $, versus N,
for $k=2,\ 3,\ 5$. The straight line
is proportional to $1/N$.}
\label{fig:phalf}
\end{figure}

To complete this argument, we have computed numerically
$p_{1/2} (N)$, i.e. the value of $p$ such that 
$\ell(N,p_{1/2}(N)) = \ell(N,0) /2 $.
Figure (\ref{fig:phalf}) shows quite clearly that, for large $N$,
$p_{1/2} (N) \sim 1/N$: a finite number of rewired
links is able to divide the mean length between vertices by two
\footnote{As shown in \cite{comment}, $N^* \sim p^{-\tau}$ implies
$p_{1/2} (N) \sim N^{-1/\tau}$; we thus have 
a clear indication that $\tau=1$.}.

\begin{figure}[bt]
\centerline{
        \epsfig{figure=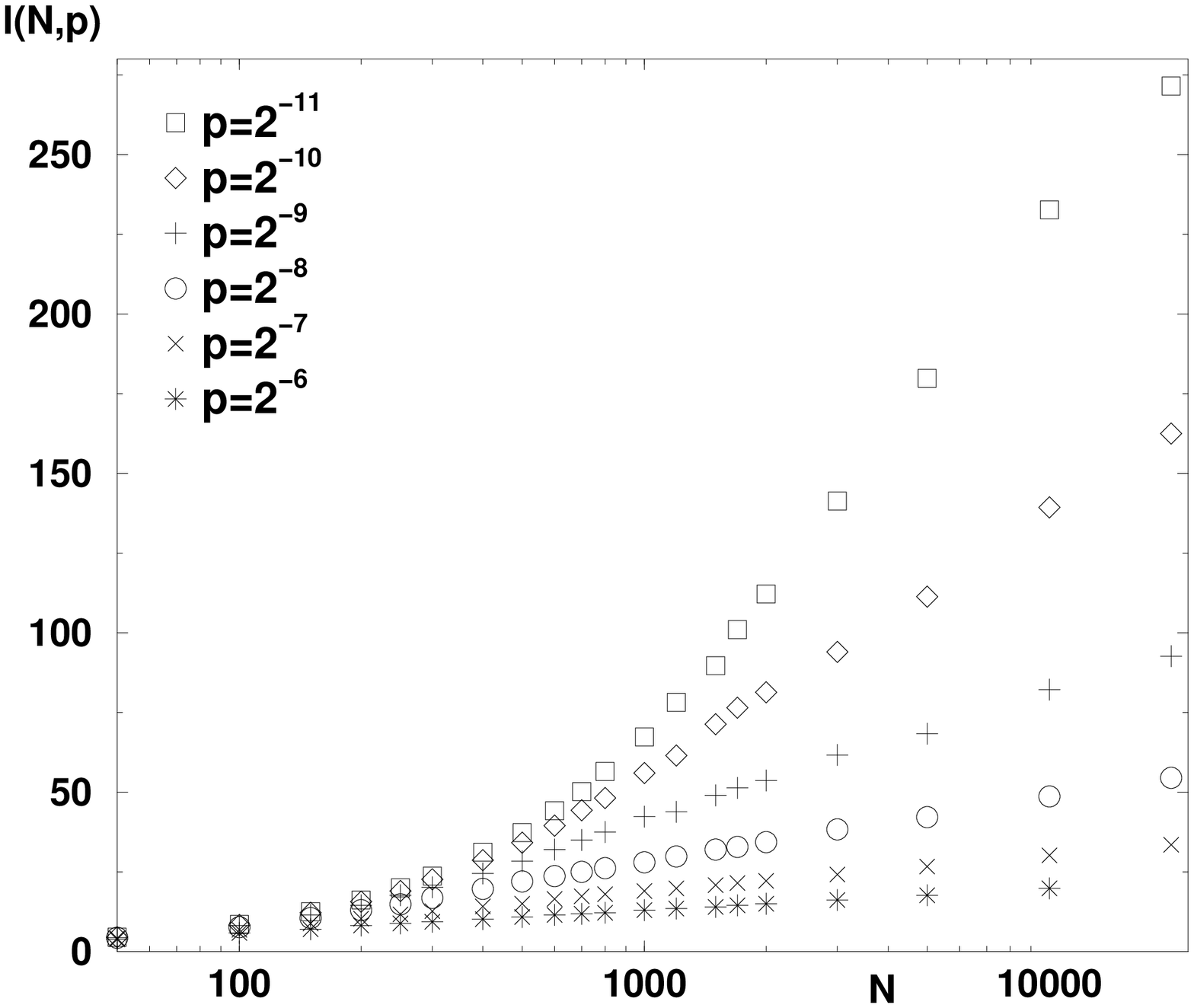,width=7cm,angle=-90}
        }
\caption{$\ell(N,p)$ versus $N$, for $p=2^{-a}$, $a=6,\cdots,11$,
and $k=3$. For large values of $p$ we have a straight line in the
semi-log plot, while for small values of $p$ we observe the
crossover between $\ell(N,p) \sim N$ and $\ell(N,p) \sim \ln(N)$.
The value of $N^*(p)$ is given by the slope of the linear part in the
semi-log plot.}
\label{fig:nstarmeasure}
\end{figure}

\begin{figure}[bt]
\centerline{
        \epsfig{figure=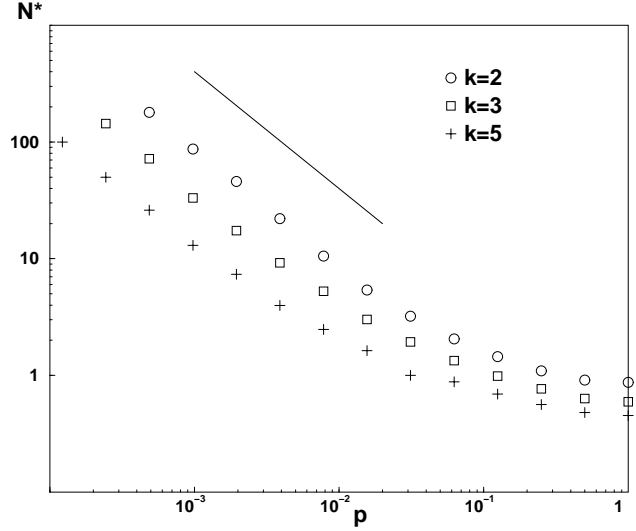,width=7cm,angle=-90}
        }
\caption{$N^*(p)$ versus $p$ for $k=2,\ 3,\ 5$. The straight line
is proportional to $1/p$.}
\label{fig:nstar}
\end{figure}

Let us now go back to the scaling hypothesis of \cite{bart}.
If the scaling form of equation (\ref{ansatz}) is valid, we have to 
compute $\ell(N,p)$ at fixed $p$ in order to estimate $N^*(p)$.
For large $N$, it behaves like $N^*(p) \ln(N)$ (see 
figure (\ref{fig:nstarmeasure}) for different values of $p$).
For small $p$, $N^*(p)$ becomes bigger and bigger, so that we have
to use larger and larger values of $N$. We show in figure 
(\ref{fig:nstar})
that the $N^*(p)$ estimated in this way behaves like $1/p$ for small
$p$ (and for $p \to 1$, $N^*(p) \to \ln(2k-1)$, in accordance
with $\ell(N,1) \sim \ln(N)/\ln(2k-1)$),
giving $\tau=1$. This is not very surprising if we consider 
the above discussion showing that a finite number of rewired links
will change the coefficient of the scaling of $\ell$ with $N$ but
not the scaling itself. Moreover, $p_{1/2} (N)$ corresponds to the 
drop in the curves of figure (\ref{fig:lmoy})
and can therefore be considered as a crossover value.

\begin{figure}[bt]
\centerline{
      \epsfig{figure=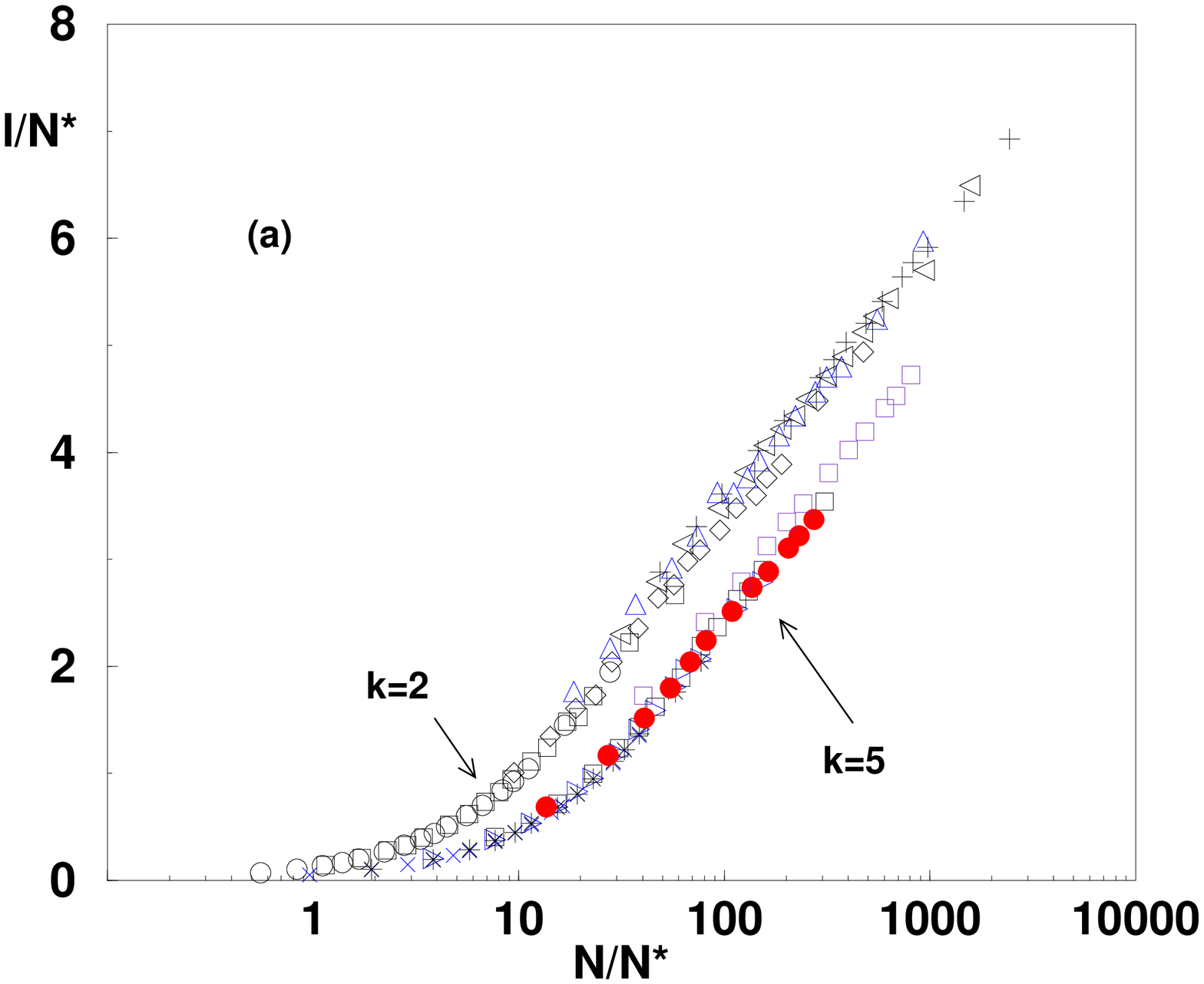,width=5cm,angle=-90}
      \epsfig{figure=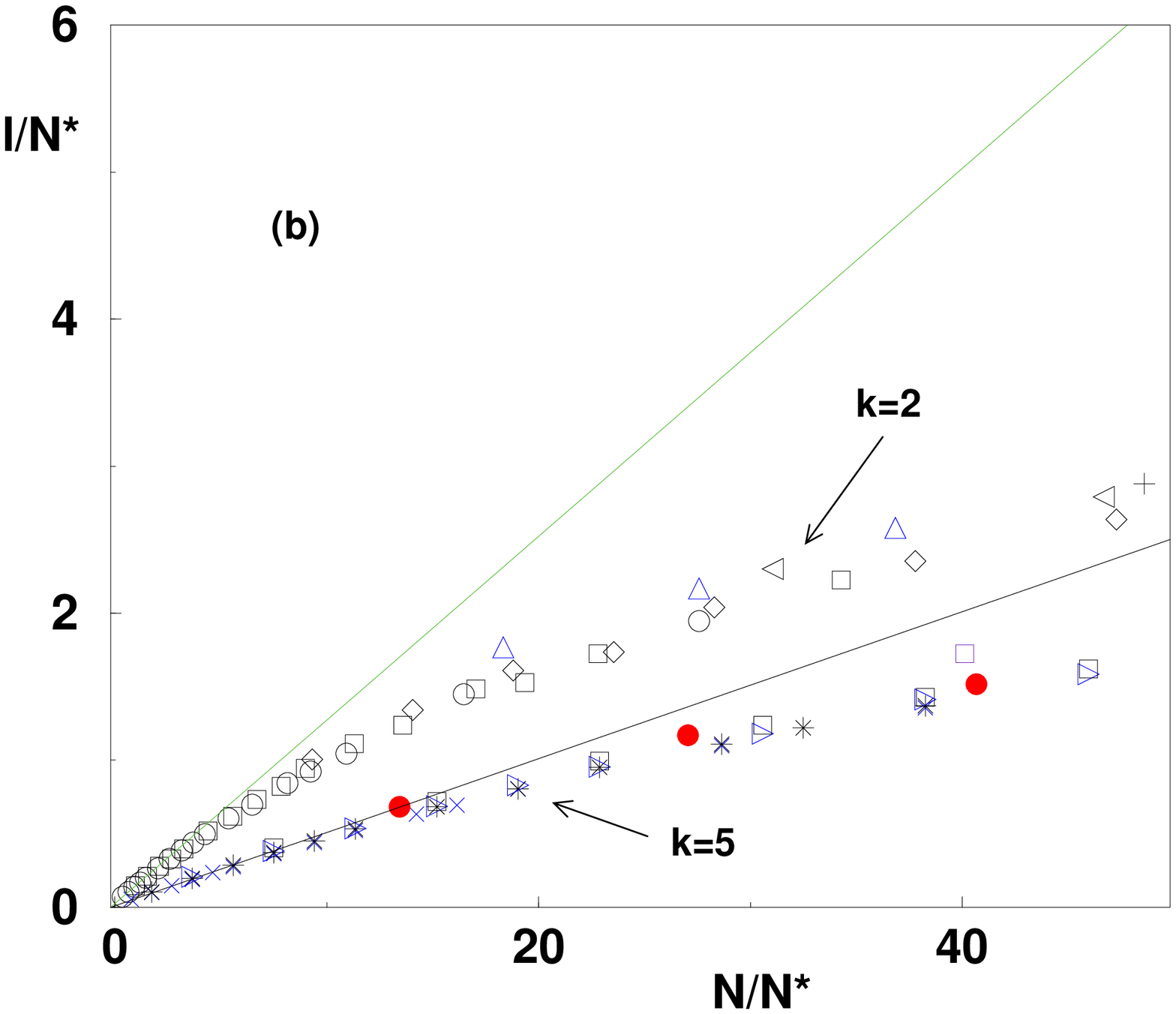,width=5cm,angle=-90}
        }
\caption{Data collapse $\ell(N,p)/N^*(p)$ versus $N/N^*(p)$, for $k=2$ and
$k=5$; (a): log-linear scale showing at large $N/N^*$ the logarithmic
behaviour; (b) linear scale showing at small $N/N^*$ the linear behaviour
$\ell(N,p) \sim N/(4k)$: the straight lines have slopes $1/8$ and $1/20$.}
\label{fig:collapse}
\end{figure}

Using the determined values of $N^*$, we plot in figure
(\ref{fig:collapse}) $\ell(N,p)/N^*(p)$ versus $N/N^*(p)$ for various values
of $N$ and $p$. We observe a nice collapse of the data for each value
of $k$. Thanks to the range of values of $N$ that we use, we are able to show
the collapse over a much wider range of values than \cite{bart}.
We clearly see the linear behaviour $F_k (x \ll  1) \sim x/(4k)$,
and the crossover to $F_k (x \gg  1) \sim \ln(x)$. 
Note that, as explained in \cite{newman}, we have to use values of $p$
lower than $1/k^2$ (and of course large
enough values of $N$, i.e. $N \gg  k$) to obtain a clean scaling behaviour: for
too large $p$, we are moving out of the scaling regime close to 
the $p=0$-transition.

\subsection{Clustering coefficient}
\label{geomC}

To define the ``small-world'' behaviour, two ingredients are used
by Watts and Strogatz \cite{watts_strogatz}. The first one is
the chemical length studied in the previous paragraph, which depends
strongly on $p$ and $N$. The second one is more local: the
``clustering coefficient'' ${\cal C}(p)$ quantifies its ``cliquishness''.
${\cal C}(p)$ is indeed defined
as follows: if $c_i$ is the number of neighbours of a vertex $i$, there
are a priori $c_i (c_i-1)/2$ possible links between these neighbours.
Denoting ${\cal C}_i$ the fraction of these links
that are really present in the graph, ${\cal C}(p)$ is the average of 
${\cal C}_i$ over all vertices. On a linear-log plot, 
${\cal C}(p)/{\cal C}(0)$ is close to $1$ for a wide range of values of $p$,
and its drop occurs around $p \approx 0.1$. This is therefore
in contrast with $\ell(N,p)$, whose drop occurs for much smaller values
of $p$ as soon as $N$ is large enough. It is therefore an interesting
question whether there is an upper threshold on $p$ for the
small-world behaviour.

We now show that a simple redefinition of ${\cal C}(p)$ leads to a very simple
formula, without altering its physical signification, nor the
shape of the curve.
For $p=0$, each vertex has $2k$ neighbours; it is easy to see that the number
of links between these neighbours is ${\cal N}_0 = 3 k (k-1)/2$. Then
${\cal C}(0)= \frac{3 (k-1)}{2 (2k-1)}$.
For $p > 0$, two neighbours of $i$ that were connected at $p=0$ are
still neighbours of $i$ and linked together with probability 
$(1-p)^3$, up to terms of order $\frac{1}{N}$. The mean number of links 
between the neighbours of a vertex is then clearly
${\cal N}_0 (1-p)^3 + O(\frac{1}{N})$.
The clustering coefficient ${\cal C}(p)$ is defined as the mean of
the ratio ${\cal C}_i = \frac{{\cal N}_i }{c_i (c_i-1)/2}$. If instead
we define $\tilde {{\cal C}} (p)$ as the ratio of the
mean number of links between the neighbours of a vertex and the
mean number of possible links between the neighbours of a vertex,
we obtain 
\begin{equation}
\tilde {{\cal C}} (p) = \frac{3(k-1)}{2(2k-1)} (1-p)^3
\label{eq:clust}
\end{equation}

\begin{figure}[bt]
\centerline{
        \epsfig{figure=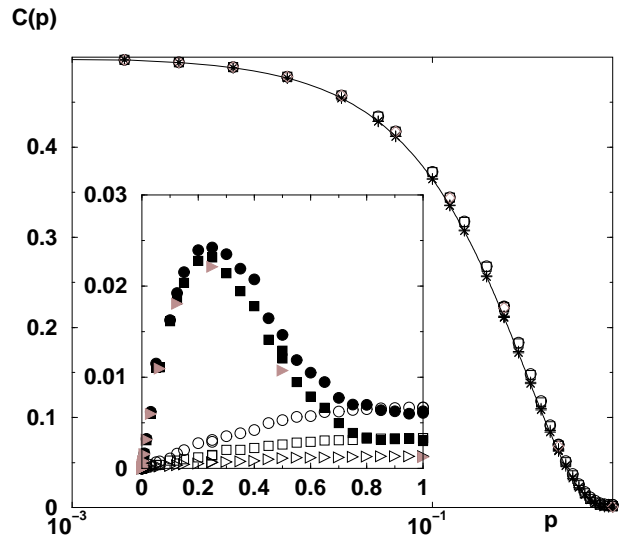,width=7cm,angle=-90}
        }
\caption{${\cal C}(p)$ and $\tilde {{\cal C}} (p)$ versus $p$,
for $k=2$ (${\cal C}(0)=\tilde {{\cal C}} (0)=0.5$),
$N=1000,\ 2000,\ 5000$: open symbols are for ${\cal C}(p)$,
and the crosses are for $\tilde {{\cal C}} (p)$; the line
is ${\cal C}(0) (1-p)^3$. Inset: corrections
${\cal C}(p)-{\cal C}(0) (1-p)^3$ (filled symbols) for $N=1000$ (circles),
$N=2000$ (squares) and $N=5000$ (triangles), and
$\tilde {{\cal C}}(p)-{\cal C}(0) (1-p)^3$ (open symbols)
for $N=1000$ (circles),
$N=2000$ (squares) and $N=5000$ (triangles). We see that the corrections
go to zero as $1/N$ for $\tilde {{\cal C}}(p)$; the
corrections for ${\cal C}(p)$ are larger, but anyway very small.
}
\label{fig:clustering}
\end{figure}

We check numerically, with $N=50$ to $N=8000$, and averaging over 
$5000$ samples, that the two definitions lead to the same
behaviour (we see in figure (\ref{fig:clustering})
that the difference between ${\cal C}(p)$ and
$\tilde {{\cal C}} (p)$ is very small), and that the corrections
to eq. (\ref{eq:clust}) are indeed of order $1/N$.
The behaviour of ${\cal C}(p)$ is therefore very simply described
by ${\cal C}(p) \approx {\cal C}(0) (1-p)^3$, and the dependence
on $N$ is very small.

To summarize this section, we have shown that the small-world
behaviour -- as defined by the average chemical distance and the
clustering coefficient -- is indeed present for any finite value of
$0<p<1$ as soon as the network is large enough. 

\section{Ising model}
\label{Ising}

In this section we want to investigate the consequences of the mixed
geometrical structure of small-world networks on an Ising model as
a prototype of
statistical-mechanics models that can be defined on it. This model can be 
understood as a continuous interpolation
of a pure one-dimensional model for $p=0$ showing no phase transition
at finite temperature to a model on a random graph\footnote{
As already mentioned in the introduction, every point
in this model has a minimal connectivity $k$. So, even in the case
$p=1$, the model is not equivalent to the usual random graph where
both endpoints of a link are chosen randomly.}
for $p=1$ 
having a finite critical temperature $T_c(p=1)>0$ as long as $k\geq2$,
cf. \cite{kanter}. 
In agreement with the results from section \ref{geom}, we find for every
finite $p>0$
that the low temperature behaviour of the model is of mean-field
character, even if we observe a finite temperature crossover to a
dominance of the one-dimensional structure. This observation confirms
the value $p_c=0$ for the onset of a non-trivial thermodynamical 
small-world behaviour as already found in the geometrical properties,
and it shows
again the crucial importance of the mixed geometrical structure, as
even global quantities can be dominated by the initial ordered
structure for high temperatures.

\subsection{General formalism}
\label{IsingA}

The system we want to study is given by its Hamiltonian
\begin{equation}
\label{hamiltonian}
H(\{S_i\}) = - \sum_{i=1}^N S_i \sum_{j=1}^k S_{m(i,j)}
\end{equation}
with $N$ Ising spins $S_i=\pm   1,\ i=1,...,N,$ and periodic boundary 
conditions, i.e. we identify $S_{N+1}=S_1$ etc. in
the following. The independently and identically distributed  
numbers $m(i,j)$ are drawn from the probability distribution 
\begin{equation}
\label{disorder}
P(m(i,j))= (1-p)\delta_{m(i,j),i+j} + \frac{p}{N} \sum_{l=1}^N
\delta_{m(i,j),l},\ 
\end{equation}
i.e. for $p=0$ we obtain a pure one-dimensional Ising model where
every site is connected to its $2k$ nearest neighbours by 
ferromagnetic bonds of strength 1, whereas this structure is completely
replaced by random long-range bonds for $p=1$. The number
of bonds in 
the model is given by $kN$, independently of the disorder strength
$p$. Here we consider only the case of finite probabilities
$p=O(1)$, i.e. an extensive number of links is rewired and,
according to the last section, we are therefore in the small-world regime.

In order to decide whether there exists a ferromagnetic
phase transition at finite temperature or not, we have to calculate
the free-energy density at inverse temperature
$\beta$. Due to the existence of an extensive number as well of random
as of one-dimensional links and due to the translational invariance of
the distribution (\ref{disorder}) we expect this quantity to be
self-averaging, we therefore have to determine
\begin{eqnarray}
\label{freeenergy}
f &=& - \lim_{N\to \infty} \frac{1}{\beta N} \overline{ \ln
Z}\nonumber\\
&=&  - \lim_{N\to \infty} \frac{1}{\beta N} \overline{ \ln
\sum_{\{S_i\}} e ^{-\beta H(\{S_i\})}}\ .
\end{eqnarray}
The average $\overline{(\cdot  )}$ over the disorder distribution
$P(m(i,j))$ is achieved  
with the help of the replica trick 
\begin{equation}
\overline{\ln Z} = \lim_{n\to 0} \partial_n \overline{Z^n}
\end{equation}
by introducing at first a positive integer number $n$ of replicas
of the original system, averaging over the disorder and sending $n\to
0$ at the end of the calculations. Thus
the replicated and disorder averaged partition function can be written
as
\begin{eqnarray}
\label{Zn}
\overline{ Z^n}& =& \sum_{\{ {\bf S}_i\}} \overline{\exp\left\{ 
-\beta\sum_{a=1}^n H(\{S^a_i\})\right\}} \nonumber\\
&=& \sum_{\{ {\bf S}_i\}} \prod_{i=1}^N \prod_{j=1}^k \left(
(1-p)e ^{\beta {\bf S}_i \cdot   {\bf S}_{i+j}} + \frac{p}{N} \sum_{l=1}^N 
e ^{\beta {\bf S}_i \cdot   {\bf S}_l} \right)
\end{eqnarray}
where we introduced the replicated Ising spins ${\bf S}_i=(S_i ^1,...,S_i
^n)$. This expression can be simplified by defining the $2^n$ order
parameters \cite{Mo} 
\begin{equation}
\label{op}
c({\bf S}) := \frac{1}{N} \sum_{i=1}^N \delta_{{\bf S}_i,{\bf S}}
\end{equation}
giving the fraction of $n$-tuples in $\{ {\bf S}_i \}$ which are equal
to ${\bf S}\in \{-1,+1\}^n$, and their conjugates $\hat{c}({\bf S})$. 
These order parameters have to be
normalized, $\sum_{{\bf S}}c({\bf S})=1$. After a change $\hat{c}\to
i\hat{c}$ leading to real order parameters, we arrive at
\begin{eqnarray}
\label{Zn2}
\overline{ Z^n}&=&\int \prod_{{\bf S}} dc({\bf S})\ d\hat{c}({\bf S})
\exp\left\{ N \left( -\sum_{{\bf S}} c({\bf S}) \hat{c}({\bf S})
+ \frac{1}{N} \ln \mbox{tr } {\bf T}^{\frac{N}{k}} \right)\right\}
\nonumber\\
&=& \int \prod_{{\bf S}} dc({\bf S})\ d\hat{c}({\bf S})
\exp\left\{ N f_n[c,\hat{c}] \right\}
\end{eqnarray}
with an effective $2^{kn}\times   2^{kn}$-transfer matrix ${\bf T}$ given by 
its entries 
\begin{equation}
\label{T}
{\bf T}({\bf S}_1,...,{\bf S}_k|{\bf S}_{k+1},...,{\bf S}_{2k})
=\prod_{i=1}^k e ^{\hat{c}({\bf S}_i)} \prod_{j=1}^k \left(
(1-p)e ^{\beta {\bf S}_i \cdot   {\bf S}_{i+j}} + p \sum_{{\bf S}} c({\bf S}) 
e ^{\beta {\bf S}_i \cdot   {\bf S}} \right)\ .
\end{equation}
At this point we remark that the small-world Ising model offers an
interesting interplay between technical concepts of mean-field
theory, as represented by the global order parameters, and the
theory of one-dimensional systems, here represented by the
effective transfer matrix. 
As in the conventional transfer matrix method, the contribution
of the second term in $f_n$  can be
determined by the largest eigenvalue of ${\bf T}$ with right (left)
eigenvector $|\lambda_r\rangle \ (\langle\lambda_l|)$, 
\begin{equation}
f_n[c,\hat{c}] = -\sum_{{\bf S}} c({\bf S}) \hat{c}({\bf S})
+\ln\frac{\langle\lambda_l|{\bf T}|\lambda_r\rangle }{
\langle\lambda_l|\lambda_r\rangle}\ ,
\end{equation}
but in order to calculate the integrals over the order parameters in
(\ref{Zn2}) we have to use the saddle point method which implies
\begin{equation}
c({\bf S}) = \sum_{{\bf S}_1,...,{\bf S}_{k-1}} \frac{
\langle\lambda_l|{\bf S},{\bf S}_1,...,{\bf S}_{k-1}\rangle 
\langle{\bf S},{\bf S}_1,...,{\bf S}_{k-1}|\lambda_r\rangle
}{\langle\lambda_l|\lambda_r\rangle}\ ,
\end{equation}
i.e. the explicit form of the transfer matrix itself depends on the
eigenvectors, and the linear structure of the eigenvalue equations is
destroyed. 

\subsection{High-temperature solution}
\label{IsingB}

The problem simplifies significantly in its high-temperature phase
where the correct solution of the saddle point equations
\begin{eqnarray}
\label{saddle}
c({\bf S}) &=& \frac{1}{N} \frac{\partial}{\partial \hat{c}({\bf S})}
\ln \mbox{tr}  {\bf T}^{\frac{N}{k}} \nonumber\\
\hat{c}({\bf S}) &=& \frac{1}{N} \frac{\partial}{\partial c({\bf S})}
\ln \mbox{tr}  {\bf T}^{\frac{N}{k}}
\end{eqnarray}
can be found without knowing the above-mentioned eigenvectors and
is given by the paramagnetic values $c_{pm}({\bf S})=1/2^n$ and 
$\hat{c}_{pm}({\bf S})=kpa ^n$.      
$a$ does not depend on ${\bf S}$, so it can be taken out of 
$\ln \mbox{tr}  {\bf T}^{\frac{N}{k}}$ and cancels finally with
$-\sum c \hat{c}$ in (\ref{Zn2}). 
In this phase all replicated spins
 ${\bf S}$ have the same density, and thus the average
magnetization $m=\lim_{n\to 0} \sum_{{\bf S}} S^1 c({\bf S})$
as well as the overlaps 
$q^{ab}=\lim_{n\to 0} \sum_{{\bf S}} S^a S^b c({\bf S})$
vanish. 

Even if this solution exists for all temperatures, it is not stable
for low temperatures. The critical temperature can be determined
by investigating the $2^{n+1}$-dimensional fluctuation matrix
\begin{equation}
\label{fluctuation}
\left(
\begin{array}{ccc}
\underline{\partial^2 f_n[c,\hat{c}]}  &\ \ &
\underline{\partial^2 f_n[c,\hat{c}]} \\
{\partial c\  \partial c}&&{\partial c\  \partial \hat{c}} \\ \\
\underline{\partial^2 f_n[c,\hat{c}]}& &
\underline{\partial^2 f_n[c,\hat{c}]}\\
{\partial \hat{c}\  \partial c}&&{\partial \hat{c}\ \partial\hat{c}}
\end{array}
\right)\ .
\end{equation}
The paramagnetic solution is valid as 
long as none of the eigenvalues of this matrix changes sign 
\footnote{Due to the common change $\hat{c}\to i\hat{c}$ one half
of the eigenvalues has to be negative, the other half positive in
order to insure a stable saddle point.}.
The phase 
transition therefore appears at the point where the first eigenvalue
becomes zero and the system becomes unstable with respect
to Gaussian fluctuations around the given saddle point. 

\subsection{Crossover from one-dimensional to mean-field behaviour}
\label{IsingC}

The problem in calculating these eigenvalues consists in the fact that
the transfer matrix ${\bf T}$ is
given by a sum over non-commuting matrices. So it is not clear how
to obtain the eigenvectors of ${\bf T}$ even at the paramagnetic saddle
point where the problem can be linearized again because we 
already know $c$
and $\hat{c}$ and the form of the transfer matrix is fixed.

At this moment we therefore restrict to the most interesting case of
small $p\ll  1$ and treat the problem by means of a 
first order perturbation theory in $p$ around the pure one-dimensional
model. In this case we are in principle able to calculate all the 
($k$-dependent) eigenvectors, which are simple direct products of $n$ 
eigenvectors
of the pure and unreplicated transfer matrix, and hence the 
perturbation-theoretic corrections to their eigenvalues. The linearized 
transfer matrix reads
\begin{eqnarray}
{\bf T}_{lin}({\bf S}_1,...,{\bf S}_k|{\bf S}_{k+1},...,{\bf S}_{2k}) &=& 
\exp\left\{\sum_{i+1}^k\hat{c}({\bf S}_i)+\beta\sum_{i,j=1}^k{\bf S}_i 
{\bf S}_{i+j} \right\} 
\nonumber \\
&&\times \left[ 1-k^2 p + p\sum_{{\bf S}} c({\bf S})
\sum_{p,q=1}^k \exp\left\{\beta {\bf S}_p ({\bf S}-{\bf S}_{p+q})
\right\} \right]\ .
\end{eqnarray}
As we show in some detail in Appendix A from the analysis of the
entries of the fluctuation matrix (\ref{fluctuation}), this
perturbation expansion contains powers of a term proportional to
$p\xi_0$ with $\xi_0$ being the correlation length of the pure system,
and its first order approximation consequently breaks down 
when $p\xi_0$ becomes larger than
$O(1)$ for increasing disorder $p$ or decreasing temperature $T$.
In the pure model the
correlation length diverges for low temperatures as
\begin{equation}
\xi_0 \propto e ^{k(k+1)\beta}\ .
\end{equation}
Consequently, at fixed but low temperature $T$, we find a crossover 
from a weakly perturbated one-dimensional behaviour for disorder 
strengths $p\ll  p_{co}(T)$ with 
\begin{equation}
p_{co}(T) \propto \exp\left\{ \frac{-k(k+1)}{T} \right\}
\end{equation}
to a disorder-dominated and hence mean-field like regime for larger
$p$. This can be understood by a simple physical argument.
We consider a cluster of correlated spins in the pure model which has
a typical length scale $l\approx \xi_0$. Thus the number of links in
this cluster is also $O(\xi_0)$ for finite $k$, and the average
number of redirected links in this cluster at disorder strength $p$
is approximately $p\xi_0$. For $p\ll  p_{co}(T)$ there are on average 
consequently less than
one redirected link per cluster, and the system is not seriously
perturbated by the disorder. The opposite holds for larger $p$.

This shows that an arbitrarily small, but finite fraction $p$ of
redirected links (``short cuts'' in the graph) leads at sufficiently
small temperature $T<T_{co}(p)$,
\begin{equation}
\label{Tco}
T_{co}(p) \propto - \frac{k(k+1)}{\log (p)},\ \ p\ll   1,
\end{equation}
to a change of the behaviour of the model from a 
one-dimensional to a mean-field one, which nicely underlines the
importance of both geometrical structures in the small-world
lattice.

\subsection{The ferromagnetic phase transition}
\label{IsingD}

\begin{figure}[bt]
\centerline{
        \epsfig{figure=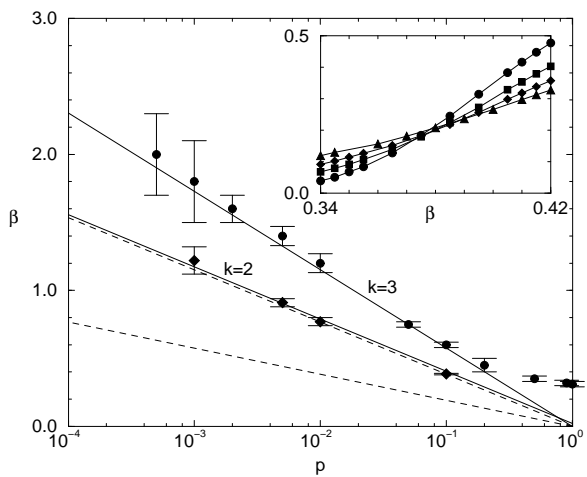,width=9cm}
        }
\caption{Inverse critical temperature $\beta_c(p)$ for $k=2$ (circles)
and $k=3$ (diamonds). The full lines show the asymptotic scaling
(\ref{Tcapprox}) of $\beta_c(p\ll  1)$. The scaling (\ref{Tco}) of the
crossover between one-dimensional and mean-field behaviour is given by
the dashed lines -- and consistently found to be at higher temperatures as the
ferromagnetic phase transition. The inset shows the $\beta$-dependence
of the Binder cumulant used to determine the critical temperature
for $p=0.1,\ k=3$ and $N=500,1000,2000,5000$ (triangles, diamonds,
squares, circles).}
\label{fig:betac_num}
\end{figure}

In the low-temperature regime $T\ll   T_{co}(p)$ the thermodynamic
behaviour is dominated by the mean-field type disorder, and we expect a
finite temperature transition to a ferromagnetically ordered phase
at finite temperature $T_c(p)$ at least for sufficiently large $p$
and $k\ge 2$. Due to the above-mentioned technical problems in
diagonalizing the transfer matrix we cannot calculate this transition
analytically, and we compute therefore the full line $T_c(p)$ for 
$k=2$ and $k=3$ by means of numerical simulations. 
We use a cluster algorithm
\cite{cluster} to compute the equilibrium distribution
of the magnetization, for system sizes
ranging from $N=500$ to $N=8000$, and use Binder cumulants 
\cite{binder} to determine
the critical point (see the inset of figure (\ref{fig:betac_num}) for 
an example).

The important result is that we obtain a transition at a 
non-zero temperature
for all the investigated values of $p$. Moreover, for small $p$ we have,
as shown in figure (\ref{fig:betac_num}):
\begin{equation}
\label{Tcapprox}
T_c(p) \propto - \frac{2 k}{\log (p)}  .
\end{equation}

This transition line is found to be always at smaller temperatures than
the crossover temperatures, which illustrates again the mean-field 
character of the phase transition. 

Even if the behaviour of the system is dominated by the random part of
its Hamiltonian, the underlying one-dimensional structure is crucial
for the existence of the phase transition and for the explicit value
of the transition temperature. This becomes clear from the fact that
only the existence of the short-range links leads to the existence of
a macroscopic cluster for $p$ below the percolation threshold of the
random bonds, and can be supported analytically by
investigating a version of the model where all one-dimensional bonds
are deleted and only the random bonds for fixed $p$ are conserved. This
model shows a ferromagnetic transition only above                 
$p_c(k)=1-\sqrt{(k-1)/k} $. So, even if the 
phase transition is induced by the presence of long-range
interactions, it is based on an interplay between both structures.

\subsection{A simplified model}
\label{IsingE}

In this subsection we present a slightly modified model where the
full procedure introduced in section \ref{IsingA} can be followed
analytically, and the phase diagram can be calculated explicitly.
The model has the same Hamiltonian (\ref{hamiltonian}), but its 
disorder distribution is given by
\begin{equation}
\tilde{P}(m(i,j))= (1-p)\delta_{m(i,j),i+1} + \frac{p}{N} \sum_{l=1}^N
\delta_{m(i,j),l}\ .
\end{equation}
So the underlying one-dimensional graph is changed: instead of having
bonds to the next $2k$ neighbours it includes $k$ bonds to each of the
two next
nearest neighbours (which, in the pure case, is equivalent to one bond
of strength $k$). In the disordered version every of these bonds is
replaced with probability $p$ by a random bond, so the random
structure of the model remains unchanged compared to the original
model. Anyway, this model remains a ``valid'' small-world network as
it consists of a mixture of a regular low-dimensional with a random
long-ranged lattice. This can e.g. be confirmed by the fact that
our simplified model also shows the scaling behaviour (\ref{ansatz})
with the same scaling exponent $\tau=1$ as the latter depends only on
the dimensionality of the regular structure, cf. \cite{newman}.
Because of the geometrical similarity of the underlying 
networks we expect also a qualitatively similar thermodynamic behaviour.

Again we average the replicated partition function over the disorder
and introduce the order parameters $c({\bf S})$ and $\hat{c}({\bf S})$.
By doing this we arrive again at
\begin{equation}
\label{Zn3}
\overline{ Z^n}
= \int \prod_{{\bf S}} dc({\bf S})\ d\hat{c}({\bf S})
\exp\left\{ N f_n[c,\hat{c}] \right\}
\end{equation}
with a slightly changed $f_n$,
\begin{equation}
\label{fn}
f_n[c,\hat{c}] = -\sum_{{\bf S}} c({\bf S}) \hat{c}({\bf S})
+ \frac{1}{N} \ln \mbox{tr} {\bf T}^N
\end{equation}
where the effective transfer matrix is of dimension $2^n$ and reads
\begin{equation}
\label{tmatrix}
{\bf T}({\bf S}_1|{\bf S}_2) = e ^{\hat{c}({\bf S}_1)}
\left[ (1-p) \exp\{ \beta{\bf S}_1\cdot   {\bf S}_2 \} +p \sum_{{\bf S}} 
c({\bf S}) \exp\{ \beta{\bf S}_1\cdot   {\bf S} \} \right]^k\ .
\end{equation}

Also in this case, the simple paramagnetic saddle point for $c$ and
$\hat{c}$ is given by $c_{pm}({\bf S})=1/2^n,\ 
\hat{c}_{pm}({\bf S})=kp a^n $ with a $\beta$-dependent $a$ canceling 
in (\ref{fn}), which therefore becomes
\begin{equation}
\label{fn2}
f_n[c_{pm},\hat{c}_{pm}] = \frac{1}{N} \ln \mbox{tr} {\bf T}^N_{pm}
\end{equation}
with
\begin{equation}
{\bf T}_{pm}({\bf S}_1|{\bf S}_2) =\left[ (1-p) \exp\{ 
\beta{\bf S}_1\cdot   {\bf S}_2 \} +p (\cosh\beta)^n \right]^k \ .
\end{equation}
This matrix can be easily diagonalized by introducing the
two-dimensional orthonormalized vectors $|+\rangle = 1/\sqrt{2}\
(1,1)$ and $|-\rangle = 1/\sqrt{2}\ (1,-1)$. The eigenvectors of
${\bf T}_{pm}$ are 
$|{\mbox{\boldmath $\mu  $}}\rangle = 
|\mu   ^1\rangle\otimes\cdots\otimes|\mu   ^n\rangle$
with $\mu   ^a=+,-$ for all $a=1,...,n$. With 
$\rho({\mbox{\boldmath $\mu  $}})$ being the
number of factors $|+\rangle$ in $|{\mbox{\boldmath $\mu  $}}\rangle$, 
the eigenvalues are found to be
\begin{equation}
\lambda[{\mbox{\boldmath $\mu  $}}] =
\lambda(\rho({\mbox{\boldmath $\mu  $}}))=
\sum_{j=0}^k {k\choose j} (p\cosh\beta^{n})^j (1-p)^{k-j}
(2\cosh(k-j)\beta)^{\rho({\mbox{\boldmath $\mu  $}})}
(2\sinh(k-j)\beta)^{n-\rho({\mbox{\boldmath $\mu  $}})}\ .
\end{equation}
The behaviour of $f_n$ in the thermodynamic limit $N\to\infty$ is
completely determined by the largest eigenvalue
 $\lambda(n)=\lambda[+...+]$,
and the paramagnetic free energy of the model reads
\begin{eqnarray}
-\beta f_{pm} &=& \lim_{n\to 0} \partial_n f_n[c_{pm},\hat{c}_{pm}]
\nonumber\\
&=& \sum_{j=0}^k {k\choose j} p^j (1-p)^{k-j} \left(
j\ln\cosh\beta +\ln2\cosh (k-j)\beta \right)\ .
\end{eqnarray}

\begin{figure}[bt]
\centerline{
        \epsfig{figure=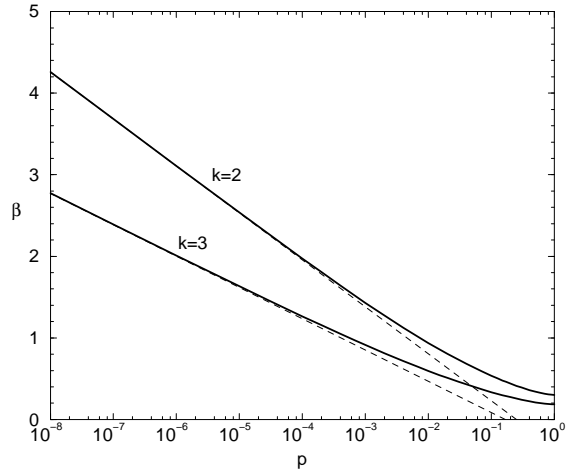,width=9cm}
        }
\caption{The inverse phase transition temperature $\beta_c(p)$ in the
simplified model for $k=2$ and $k=3$. The dashed lines show the
asymptotic behaviour given in (\ref{temp}).}
\label{fig:betac_an}
\end{figure}

The second eigenvalue $\lambda(n-1)=\lambda[-+...+]$ of the transfer 
matrix ${\bf T}_{pm}$ describes in the replica limit $n\to 0$
the decay of the two-point correlation function $\overline{\langle S_i
S_j \rangle}\propto\lambda(n-1)^{|i-j|}$ for distances 
$1 \ll   |i-j| \ll   \ell (N,p)$, cf. section \ref{geomB},
i.e. for points $i$ and $j$ whose chemical distance is given with 
finite probability 
by the one-dimensional distance $|i-j|$ and does not include
random bonds. The corresponding correlation length reads
\begin{eqnarray}
\xi_p &=& -\lim_{n\to 0} \frac{1}{\ln\lambda(n-1)}\nonumber\\
&=& \frac{-1}{\ln\left( \sum_{j=0}^k {k\choose j} p^j (1-p)^{k-j}
\tanh (k-j)\beta \right)}
\end{eqnarray}
and remains finite for every non-zero temperature. So, in complete
agreement with our findings for the original model in the last
subsections, we can conclude that the modified model has no 
ferromagnetic phase
transition caused by a divergence of the one-dimensional 
correlation length. There is nevertheless a transition due to the
fact that the paramagnetic saddle point $c_{pm}({\bf S})$ and
$\hat{c}_{pm}({\bf S})$ becomes unstable at a certain temperature. 
In order to see this we investigate again the fluctuation matrix
(\ref{fluctuation}) for the present model. The four blocks can be
calculated (see appendix B for details), and diagonalized
simultaneously. The fluctuation mode becoming at first unstable
leads to the reduced matrix
\begin{equation}
\label{fluc}
\left(\begin{array}{cc}
\Lambda_{cc} & \Lambda_{c\hat{c}}\\
\Lambda_{c\hat{c}} & \Lambda_{\hat{c}\hat{c}}
\end{array}
\right)
\end{equation}
with entries
\begin{eqnarray}
\Lambda_{cc} &=& k(k-1)p^2(\tanh\beta)^2 + 
\frac{2k^2p^2 \sum_{m=0}^{k-1} {k-1 \choose m} p^m
(1-p)^{k-m-1} \left[ \tanh (k-m-1)\beta \ (\tanh\beta)^2 
\right]}{ 1- \sum_{m=0}^{k} {k \choose m} p^m
(1-p)^{k-m} \tanh (k-m)\beta}\nonumber\\
\Lambda_{c\hat{c}} &=& -1+ \frac{kp\tanh\beta\left[1+\sum_{m=0}^{k-1} 
{k-1 \choose m} p^m
(1-p)^{k-m-1} \tanh (k-m-1)\beta\right]}{ 1-\sum_{m=0}^{k} 
{k \choose m} p^m (1-p)^{k-m} \tanh (k-m)\beta}\nonumber\\
\Lambda_{\hat{c}\hat{c}} &=& 1+2\frac{\sum_{m=0}^{k} 
{k \choose m} p^m (1-p)^{k-m} \tanh (k-m)\beta}{1-\sum_{m=0}^{k} 
{k \choose m} p^m (1-p)^{k-m} \tanh (k-m)\beta}.
\end{eqnarray}
The vanishing of its determinant gives  the critical temperature 
$T_c(p)$ which
depends on $p$. The determinant is negative for $p=0$ at all positive
temperatures, where the paramagnetic solution is known to be correct, 
and positive at $T=0$ for all $p>0$, we thus
conclude that $T_c(p>0)>0$. The explicit value can be calculated 
numerically
from (\ref{fluc}) and is shown in figure (\ref{fig:betac_an}).
The critical temperature for small disorder $p$ behaves like
\begin{equation}
\label{temp}
T_c(p) \approx - \frac{2 k}{\log (2kp)}\ ,
\end{equation}
it consequently shows the same asymptotic $p$-dependence as in the
original model, cf. (\ref{Tcapprox}). In addition it shows in this
case the   same $p$-dependence as the crossover 
temperature found
from $2kp\xi_0\propto 1$ with 
$\xi_0 =-1/\ln( \tanh k\beta ) \propto \exp\{2k\beta\}$ for $\beta\gg   1$.

\section{Summary and conclusion}

In conclusion, in the first part of this work
we have studied the geometrical properties of
small-world networks which interpolate continuously between a 
one-dimensional ring and a certain random graph. The coexistence
of a more and more diluted local structure  and of random long-ranged
links leads to some very interesting features:
\begin{itemize}
\item Due to the local structure two neighbouring vertices have
in general common neighbours, a fact which leads to a certain cliquishness.
The clustering coefficient, measuring this property,
was found to decrease like $(1-p)^3$ with the fraction $p$ of
randomly rewired links.
\item The average length between two points characterizing global
properties of the network was found to depend strongly on the amount
of disorder in the network. A crossover, first proposed in
\cite{bart}, could be  worked out: At fixed $p$, the average length
between two vertices
was found to grow linearly with the system size $N\ll  O(1/p)$
for small networks,
whereas it grows only logarithmically for large networks $N\gg  O(1/p)$. 
\end{itemize}
Therefore, the mere notion of ``small-world'' graph, i.e. the region
of disorder where the local properties are still similar to those of
the one-dimensional ring whereas the global properties are determined 
by the random short-cuts in the graph,  depends on its size, and can be 
extended to smaller and smaller $p$, taking larger and larger $N$.

In the second part these findings where corroborated by the
investigation of an Ising model defined on the small-world network.
In the thermodynamic limit we found the following behaviour for
fixed disorder strength $p$: for large temperatures, the system 
behaves very similarly to the pure one-dimensional system, whereas it
undergoes a crossover to a mean-field like region for smaller
temperatures. Finally, at low but non-zero temperature, we find 
a ferromagnetic phase transition. This underlines again the results
of the geometrical investigations that the graph is in its
small-world regime for any disorder strength at sufficiently large 
system sizes, i.e. in a region where both geometrical structures
lead to interesting physical effects.

{\bf Acknowledgment}: We are very grateful to G. Biroli, R. Monasson
and R. Zecchina for numerous fruitful discussions. MW acknowledges
financial support by the German Academic Exchange Service (DAAD).

\appendix

\section{Breakdown of the first order perturbation theory}

In this appendix we want to present the first-order perturbation
calculations for small disorder strengths $p\ll  1$ leading finally
to the crossover phenomenon described in section \ref{IsingC}. We start
from the linearized transfer matrix
\begin{eqnarray}
{\bf T}_{lin}({\bf S}_1,...,{\bf S}_k|{\bf S}_{k+1},...,{\bf S}_{2k}) &=& 
\exp\left\{\sum_{i+1}^k\hat{c}({\bf S}_i)+\beta\sum_{i,j=1}^k{\bf S}_i 
{\bf S}_{i+j} \right\} 
\nonumber \\
&&\times \left[ 1-k^2 p + p\sum_{{\bf S}} c({\bf S})
\sum_{p,q=1}^k \exp\left\{\beta {\bf S}_p ({\bf S}-{\bf S}_{p+q})
\right\} \right]\ .
\end{eqnarray}
and calculate the elements of the fluctuation matrix
(\ref{fluctuation}) around the paramagnetic saddle point up to first 
order in $p$. In order to achieve this we use the  $2^k$ 
(bi-)orthonormalized eigenvectors
$|\lambda_\alpha\rangle,\ (\langle\lambda_\alpha|)\ \alpha=1,...,2^k$,
of the pure and unreplicated transfer matrix
\begin{equation}
{\bf T}^{(0)}({S}_1,...,{S}_k|{S}_{k+1},...,{S}_{2k}) =
\exp\left\{\beta\sum_{i,j=1}^k{S}_i {S}_{i+j} \right\}\ .
\end{equation}
We choose these eigenvectors to be ordered according to their
eigenvalues.
The eigenvectors of the replicated pure system are therefore given
by $|{\mbox{\boldmath $\alpha$}}\rangle = |\lambda_{\alpha ^1}\rangle
\otimes\cdots\otimes|\lambda_{\alpha ^n}\rangle$, and the corrections
of $O(p)$ can be calculated by using these vectors.

At first we realize that the second derivative of
\begin{equation}
f_n[c,\hat{c}] = -\sum_{{\bf S}} c({\bf S}) \hat{c}({\bf S})
+\frac{1}{N} \ln \mbox{tr {\bf T}}_{lin}^{\frac{N}{k}}
\end{equation}
with respect to $c$
is already of order $p^2$ and can therefore be neglected. The
interesting entries of the fluctuation matrix consequently come from
the off-diagonal blocks $\partial^2f_n/\partial c\partial\hat{c}$.
We calculate the derivatives
\begin{eqnarray}
\label{2ndderiv}
\frac{\partial f_n}{\partial \hat{c}({\bf S})} &=&
-c({\bf S})+\frac{\sum_{{\bf S}_1,...,{\bf S}_{k-1}} {\bf
T}_{lin}^{\frac{N}{k}}({\bf S},{\bf S}_1,...,{\bf S}_{k-1}|
{\bf S},{\bf S}_1,...,{\bf S}_{k-1})}{
\mbox{tr {\bf T}}_{lin}^{\frac{N}{k}}}\\
\frac{\partial^2 f_n}{\partial \hat{c}({\bf S})\partial c({\bf R})}
&=& -\delta_{{\bf S},{\bf R}} - \frac{N}{k}c_{pm}\hat{c}_{pm}
\nonumber\\
&&+ \sum_{j=0}^{N-1}
\frac{\sum_{{\bf S}_1,...,{\bf S}_{k-1}}\left({\bf T}_{lin}^j 
\frac{\partial {\bf T}_{lin}}{\partial c({\bf R})} 
{\bf T}_{lin}^{N-j-1}
\right)({\bf S},{\bf S}_1,...,{\bf S}_{k-1}|
{\bf S},{\bf S}_1,...,{\bf S}_{k-1})}{ 
\mbox{tr {\bf T}}_{lin}^{\frac{N}{k}}} \nonumber\ .
\end{eqnarray} 
Due to the fact that
\begin{equation}
\frac{\partial {\bf T}}{\partial c({\bf R})} =p 
\sum_{p,q=1}^k \exp\left\{\beta {\bf S}_p ({\bf S}-{\bf S}_{p+q})
+\beta\sum_{i,j=1}^k {\bf S}_i {\bf S}_{i+j}
+\sum_{i=1}^k \hat{c}({\bf S}_i) \right\}
\end{equation}
is already linear in $p$, the other ${\bf T}_{lin}$-factors can
be replaced by the replication $({\bf T}^{(0)})^{\otimes n}$
of the pure matrix. Introducing two-times the identity
\begin{equation}
{\bf 1}=\sum_{{\mbox{\boldmath $\alpha$}}} |{\mbox{\boldmath
$\alpha$}}\rangle\langle {\mbox{\boldmath $\alpha$}}|
\end{equation}
where $\langle {\mbox{\boldmath $\alpha$}}|$ denotes the biorthogonal
set of left eigenvectors into (\ref{2ndderiv}) and keeping only the
exponentially dominant terms proportional to $\lambda_1^{nN}$, we
can write
\begin{equation}
\frac{\partial^2 f_n}{\partial \hat{c}({\bf S})\partial c({\bf R})}
= -\delta_{{\bf S},{\bf R}}- \frac{N}{k}c_{pm}\hat{c}_{pm}
+{\bf M}_{(1...1)}({\bf S},{\bf R})
+\sum_{{\mbox{\boldmath $\alpha$}}\neq(1...1)} \frac{p}{1-
\lambda_{{\mbox{\boldmath $\alpha$}}}} 
{\bf M}_{{\mbox{\boldmath $\alpha$}}}({\bf S},{\bf R})
\end{equation} 
and in the limit $n\to 0$ the fluctuation modes respecting the 
normalization of $c({\bf S})$ give rise to eigenvectors of the
form 
\begin{equation}
-1+\frac{p}{1-\lambda_2/\lambda_1} O(p^0, (e ^\beta)^0)+...
\end{equation}
with $\lambda_1$ and $\lambda_2$ being the two largest eigenvectors of
${\bf T}^{(0)}$. For low temperatures, where $1-\lambda_2/\lambda_1\ll 
1$, we have
\begin{equation}
\frac{1}{1-\lambda_2/\lambda_1}=\frac{1}{1-\exp(-\frac{1}{\xi_0})}
\approx \xi_0
\end{equation}
and the correction in $O(p)$ gets arbitrarily large for low enough
temperatures $T$. This leads directly to the crossover in the
behaviour of the model for $p\propto \xi_0^{-1}$ discussed in section
\ref{IsingC}.

\section{Fluctuations around the paramagnetic saddle point}

In this appendix we are going to present the calculations of the
Gaussian fluctuation matrix at the paramagnetic saddle point
solution for the modified model presented in Section IV.E in order 
to determine the ferromagnetic phase transition temperature for
general $k$ and $p$. We start with equations
(\ref{fn},\ref{tmatrix}),
\begin{equation}
f_n[c,\hat{c}] = -\sum_{{\bf S}} c({\bf S}) \hat{c}({\bf S})
+ \frac{1}{N} \ln \mbox{tr} {\bf T}^N\ ,
\end{equation}
\begin{equation}
{\bf T}({\bf S}_1|{\bf S}_2) = e ^{\hat{c}({\bf S}_1)}
\left[(1-p)\exp\{ \beta{\bf S}_1\cdot   {\bf S}_2\} 
+p \sum_{{\bf S}} c({\bf S}) \exp\{ \beta{\bf S}_1\cdot   {\bf S}\} 
 \right]^k\ .
\end{equation}

In the following we need the first and second derivatives of ${\bf T}$:
\begin{eqnarray}
\frac{\partial {\bf T}({\bf S}_1|{\bf S}_2)}{\partial c({\bf S})}
&=& kp\exp\{\hat{c}({\bf S}_1)+\beta {\bf S}_1\cdot   {\bf S}\} 
\left[(1-p)\exp\{ \beta{\bf S}_1\cdot   {\bf S}_2\} 
+p \sum_{{\bf S}} c({\bf S}) \exp\{ \beta{\bf S}_1\cdot   {\bf S}\} 
 \right]^{k-1}\nonumber\\
\frac{\partial {\bf T}({\bf S}_1|{\bf S}_2)}{\partial \hat{c}({\bf S})}
&=& {\bf T}({\bf S}_1|{\bf S}_2) \delta_{{\bf S}_1,{\bf S}}\nonumber\\
\frac{\partial^2 {\bf T}({\bf S}_1|{\bf S}_2)}{\partial c({\bf S})
\partial c({\bf R})} &=& k(k-1)p^2 \exp\{\hat{c}({\bf S}_1)+
\beta {\bf S}_1\cdot   ({\bf S} +{\bf R}) \}
\nonumber \\
&&\times \left[(1-p)\exp\{ \beta{\bf S}_1\cdot   {\bf S}_2\} 
+p \sum_{{\bf S}} c({\bf S}) \exp\{ \beta{\bf S}_1\cdot   {\bf S}\} 
 \right]^{k-2}\nonumber\\ 
\frac{\partial^2 {\bf T}({\bf S}_1|{\bf S}_2)}{\partial c({\bf S})
\partial \hat{c}({\bf R})} &=& \frac{\partial {\bf T}({\bf S}_1|{\bf S}_2)
}{\partial c({\bf S})} \delta_{{\bf S}_1,{\bf R}}\nonumber\\
\frac{\partial^2 {\bf T}({\bf S}_1|{\bf S}_2)}{\partial \hat{c}({\bf S})
\partial \hat{c}({\bf R})} &=& {\bf T}({\bf S}_1|{\bf S}_2) 
\delta_{{\bf S}_1,{\bf S}} \delta_{{\bf S}_1,{\bf R}}
\end{eqnarray}

The resulting saddle point equations for the calculation of 
$\overline{Z^n}$,
\begin{eqnarray}
c({\bf S}) &=& \frac{{\bf T}^N({\bf S}|{\bf S})}{\mbox{tr} 
{\bf T}^N} \nonumber\\
\hat{c}({\bf S})&=& \frac{{\bf T}^{N-1}\partial_{c({\bf S})}{
\bf T}}{\mbox{tr}
{\bf T}^N}\ ,
\end{eqnarray}
have obviously a simple paramagnetic solution of the form $c({\bf S})
= 1/2^n$ and $\hat{c}({\bf S})=2p a(\beta)^n$, i.e. a solution, where
every replicated spin has equal probability. 
Whether this is correct or not for any finite temperature depends on the
eigenvalues of the Hessian matrix
\begin{equation}
\label{hessian}
\left(
\begin{array}{ccc}
\underline{\partial^2 f_n[c,\hat{c}]}  &\ \ &
\underline{\partial^2 f_n[c,\hat{c}]} \\
{\partial c\  \partial c}&&{\partial c\  \partial \hat{c}} \\ \\
\underline{\partial^2 f_n[c,\hat{c}]}& &
\underline{\partial^2 f_n[c,\hat{c}]}\\
{\partial \hat{c}\  \partial c}&&{\partial \hat{c}\ \partial\hat{c}}
\end{array}
\right)
\end{equation}
calculated at the before-mentioned saddle point.
One important observation is that
the structure of all four blocks in this matrix is the same, resulting
in the possibility of a simultaneous diagonalization of the four
blocks, so only the submatrices of 4 eigenvalues belonging to the same
eigenvectors have to be considered. But at first we have to calculate
the entries of (\ref{hessian}), and we start with the upper left
corner:
\begin{eqnarray}
\label{ul}
\frac{\partial^2 f_n}{\partial c({\bf S}) \partial c({\bf R})} &=&
-N \hat{c}({\bf S})\hat{c}({\bf R}) + 
\frac{\mbox{tr }{\bf T}^{N-1}\  \partial^2 {\bf T} / \partial 
c({\bf S}) \partial 
c({\bf R})}{ \mbox{tr }{\bf T}^N} \nonumber\\
&+& \sum_{j=0}^{N-2} \frac{
\partial {\bf T}/\partial c({\bf S})\ {\bf T}^j\ \partial {\bf T}/
\partial c({\bf R})\ 
{\bf T}^{N-j-2}}{ \mbox{tr }{\bf T}^N}\ .
\end{eqnarray}
The numerator of the second term is dominated by the largest
eigenvalue of ${\bf T}$ which, according to the notation in III.E, is
$|+...+\rangle$. We are only interested in the limit $n\to 0$, so we
can set all $n$-th powers to 1 for the simplicity of our
calculations. 
\begin{eqnarray}
\label{b7}
\mbox{tr }{\bf T}^{N-1}\frac{\partial^2 {\bf T}}{\partial 
c({\bf S}) \partial c({\bf R})} &=& \lambda(n)^{N-1} 
\langle+...+|\frac{\partial^2 {\bf T}}{
\partial c({\bf S}) \partial c({\bf R})} |+...+\rangle\nonumber\\
&=& k(k-1)p^2 \sum_{{\bf S}_1,{\bf S}_2} \exp\{\hat{c}({\bf S}_1)+
\beta {\bf S}_1\cdot   ({\bf S} +{\bf R}) \} \left[(1-p)\exp\{ 
\beta{\bf S}_1\cdot   {\bf S}_2\} +p \right]^{k-2}\nonumber\\ 
&=& k(k-1)p^2 e ^{\hat{c}} (\cosh 2\beta)^{\frac{{\bf S}\cdot 
{\bf R}}{2}}
\end{eqnarray}
The last term in equation (\ref{ul}) is exponentially dominated by
\begin{eqnarray}
\mbox{tr } \frac{\partial {\bf T}}{\partial c({\bf S})} {\bf T}^j
\frac{\partial {\bf T}}{\partial c({\bf R})} {\bf T}^{N-j-2} &=&
\lambda(n)^{N-2} \langle +...+|\frac{\partial {\bf T}}{\partial 
c({\bf S})}|+...+\rangle\langle +...+|\frac{\partial {\bf T}}{\partial 
c({\bf R})}|+...+\rangle\nonumber\\
&+& \sum_{{\mbox{\boldmath $\mu  $}}\neq(+...+)} 
\lambda[{\mbox{\boldmath $\mu  $}}]^{j} 
\lambda(n)^{N-j-2} \langle +...+|\frac{\partial {\bf T}}{\partial 
c({\bf S})}|{\mbox{\boldmath $\mu  $}}\rangle\langle 
{\mbox{\boldmath $\mu  $}}|\frac{\partial 
{\bf T}}{\partial c({\bf R})}|+...+\rangle\nonumber\\
&+& \sum_{{\bf \mu   }\neq(+...+)} 
\lambda[{\mbox{\boldmath $\mu  $}}]^{N-j-2} 
\lambda(n)^{j} \langle{\mbox{\boldmath $\mu  $}} 
|\frac{\partial {\bf T}}{\partial 
c({\bf S})}|+...+\rangle\langle +...+|\frac{\partial 
{\bf T}}{\partial c({\bf R})}|{\mbox{\boldmath $\mu  $}}\rangle\ .
\end{eqnarray}
With
\begin{eqnarray}
\langle +...+|\frac{\partial {\bf T}}{\partial c({\bf S})}
|{\mbox{\boldmath $\mu  $}}\rangle &=&
\sum_{{\bf S}_1,{\bf S}_2} kpe ^{\hat{c}+\beta {\bf S}_1\cdot   {\bf S}}
\left( (1-p)e ^{\beta {\bf S}_1\cdot   {\bf S}_2} +p\right)^{k-1}
\langle{\bf S}_2|{\mbox{\boldmath $\mu  $}}\rangle\nonumber\\
&=&  kpe ^{\hat{c}} \sum_{m=0}^{k-1} {k-1 \choose m} p^m (1-p)^{k-m-1}
\left[ \tanh (k-m-1)\beta \ \tanh\beta \right]^{n-
\rho({\mbox{\boldmath $\mu  $}})} 
\langle{\bf S}|{\mbox{\boldmath $\mu  $}}\rangle\nonumber\\
\langle {\mbox{\boldmath $\mu  $}}|\frac{\partial {\bf T}}{\partial 
c({\bf R})}|+...+\rangle &=&
\sum_{{\bf S}_1,{\bf S}_2} kpe ^{\hat{c}+\beta {\bf S}_1\cdot   {\bf }}
\left( (1-p)e ^{\beta {\bf S}_1\cdot   {\bf S}_2} +p\right)^{k-1}
\langle{\mbox{\boldmath $\mu  $}}|{\bf S}_1\rangle\nonumber\\
&=&  kpe ^{\hat{c}} \left[ \tanh\beta \right]^{n-
\rho({\mbox{\boldmath $\mu  $}})} 
\langle{\mbox{\boldmath $\mu  $}}|{\bf R}\rangle
\end{eqnarray}
we consequently find
\begin{eqnarray}
\label{b10}
\frac{\mbox{tr } \frac{\partial {\bf T}}{\partial c({\bf S})} 
{\bf T}^j \frac{\partial {\bf T}}{\partial c({\bf R})} 
{\bf T}^{N-j-2}}{\mbox{tr } {\bf T}^N} &=&
k^2p^2 \sum_{\bf \mu   }\sum_{m=0}^{k-1} {k-1 \choose m} p^m
(1-p)^{k-m-1}
\nonumber\\&&\times \left[ \tanh (k-m-1)\beta \ (\tanh\beta)^2 
\right]^{n-\rho({\mbox{\boldmath $\mu  $}})} \nonumber\\&&\times 
\left( \lambda[{\mbox{\boldmath $\mu  $}}]^j + 
\lambda[{\mbox{\boldmath $\mu  $}}]^{N-j-2} - 
\delta_{{\mbox{\boldmath $\mu  $}},(+...+)} \right)
\langle{\bf S}|{\mbox{\boldmath $\mu  $}}\rangle\ 
\langle{\mbox{\boldmath $\mu  $}}|{\bf R}\rangle\ .
\end{eqnarray}
It is now obvious that the matrix $\partial^2 f_n/\partial c \partial
c$ has also the eigenvectors $|{\bf \mu   }\rangle$. The first one,
$|+...+\rangle$, corresponds to fluctuations changing the
normalization of $c({\bf S})$ and is not allowed. So the second one,
$|-+...+\rangle$ (or any other with 
$\rho({\mbox{\boldmath $\mu  $}})=n-1$), is
expected to be the dangerous one leading finally to the ferromagnetic
phase transition in the Ising model. From (\ref{ul},\ref{b7},\ref{b10})
we obtain for this eigenvalue
\begin{equation}
k(k-1)p^2(\tanh\beta)^2 + 
\frac{2k^2p^2 \sum_{m=0}^{k-1} {k-1 \choose m} p^m
(1-p)^{k-m-1} \left[ \tanh (k-m-1)\beta \ (\tanh\beta)^2 
\right]}{ 1- \sum_{m=0}^{k} {k \choose m} p^m
(1-p)^{k-m} \tanh (k-m)\beta}\ .
\end{equation}
The calculation of the other elements of the fluctuation matrix is
done analogously. Here we report only the results. The eigenvalue of
$\partial^2 f_n/\partial c \partial \hat{c}$ corresponding to the
eigenvector $|-+...+\rangle$ is found to be
\begin{equation}
-1+ \frac{kp\tanh\beta\left[1+\sum_{m=0}^{k-1} {k-1 \choose m} p^m
(1-p)^{k-m-1} \tanh (k-m-1)\beta\right]}{ 1-\sum_{m=0}^{k} 
{k \choose m} p^m (1-p)^{k-m} \tanh (k-m)\beta}\ ,
\end{equation}
and for $\partial^2 f_n/\partial \hat{c} \partial \hat{c}$ we get the
entry 
\begin{equation}
1+2\frac{\sum_{m=0}^{k} 
{k \choose m} p^m (1-p)^{k-m} \tanh (k-m)\beta}{1-\sum_{m=0}^{k} 
{k \choose m} p^m (1-p)^{k-m} \tanh (k-m)\beta}
\end{equation}
leading to (\ref{fluc}).

\end{document}